\documentclass[prl,twocolumn,preprintnumbers,amsmath,amssymb,superscriptaddress]{revtex4}
\usepackage{graphicx}
\usepackage{amsfonts}
\usepackage{amssymb}
\usepackage{amsmath}
\usepackage{natbib}
\usepackage{color}
\usepackage{float}
\usepackage{epstopdf}
\usepackage{hyperref,cleveref}

\newcommand{\udt}[3]{#1^{#2}_{\phantom{#2}#3}}

\newcommand{\dut}[3]{#1_{#2}^{\phantom{#2}#3}}

\begin{document}
\title{On Reconstruction of Extended Teleparallel Gravity from the Cosmological Jerk Parameter}

\author{Soumya Chakrabarti\footnote{soumya@cts.iitkgp.ernet.in}}
\affiliation{Centre for Theoretical Studies,\\
Indian Institute of Technology Kharagpur, \\
Kharagpur 721 302, West Bengal, India
}

\author{Jackson Levi Said\footnote{jackson.said@um.edu.mt}}
\affiliation{Institute of Space Sciences and Astronomy, University of Malta, Msida, MSD 2080, Malta}
\affiliation{Department of Physics, University of Malta, Msida, MSD 2080, Malta}

\author{Kazuharu Bamba\footnote{bamba@sss.fukushima-u.ac.jp}}
\affiliation{Division of Human Support System, Faculty of Symbiotic Systems Science, Fukushima University,
Fukushima 960-1296, Japan.}

\date{\today}

\begin{abstract}
Teleparallel gravity offers a path to resolve a number of longstanding issues in general relativity by re-interpreting gravitation as an artifact of torsion rather than curvature. The present work deals with cosmological solutions in an extension of teleparallel gravity. A reconstruction scheme of the theory has been proposed based on the cosmological jerk parameter. The work contains analysis of ensuing cosmological parameters for different viable models and the stability of the models against cosmic time through an investigation of perturbation of matter overdensity and the hubble parameter.
\end{abstract}
\maketitle

\section{I. Introduction}
Observational Astronomy has developed notably over the course of past few decades, resulting in phenomenal discoveries such as the accelerated expansion of the universe \cite{obs}, and more recently the accurate measurement of the present time Hubble parameter \cite{Akrami:2018odb, Aghanim:2018eyx}. General Theory of Relativity (GR) has provided a very successful description of gravity so far, however, to reproduce the non-trivial acceleration of the universe, some correction or modification of GR is to be considered. For instance, under the scope of GR, it may be assumed that the universe is dominated at late times by a cosmological constant. Moreover, galaxies and their clusters ensures the existence of a large proportion of dark matter which drives them to have a stronger effect of gravitation and not break apart \cite{Peter:2012rz}. Together, these form the $\Lambda$CDM model which produces late time accelerated expansion of the universe \cite{CST} and alongwith some modifications, can produce the correct early-time behavior of the universe as well. However, the cosmological constant alternative has it's own consistency issues mainly driven by the disparity between the cosmological constant density and it's realization in the quantum regime \cite{weinberg}. Moreover, this model fails to settle the singularity issues surrounding black holes and the big bang, and has a seemingly growing tension in some state parameters at late times \cite{Aghanim:2018eyx}. A very popular avenue to mimic the accelerated expansion is to impose a time-varying scalar field \cite{scalarfield1, scalarfield2, scalarfield3, scalarfield4} in the energy-momentum distribution that may link early time dynamics with the late time accelerated expansion. A parrallely important scheme is to consider modifications of gravity where the action of the theory is modified to write a general theory. The modifying terms take the form of an effective exotic fluid or the `dark energy' that drives late time accelerating expansion of the Universe. \medskip

Teleparallel theory is an alternative theory of gravity which is equivalent to GR at the level of the field equations for particular choices of the Lagrangian. For a comprehensive review, we refer to the monographs by Aldrovandi and Pereira \cite{Aldrovandi:2013wha}, Cai, Capozziello, De-Laurentis and Saridakis \cite{caireview}. The fundamental difference is that while GR is characterized by expressing gravitation through curvature by means of a Levi-Civita connection, Teleparallel gravity uses the Weitzenbock connection to replace curvature with torsion $T$ or a generalized function $f(T)$. Such a generalization of GR was discussed by Hehl, Von Der Heyde, Kerlick and Nesterin \cite{telep1} with the spin of matter and it's mass playing the dynamical role. Hayashi and Shirafuji \cite{telep2} formulated the theory based on the Weitzenböck space-time and characterized by a vanishing curvature tensor, dubbed as the 'new general relativity'. Generalization of the theory, named as the Einstein-Hayashi-Shirafuji theory, was examined by Flanagan and Rosenthal \cite{telep3}. This class of theories were actually first proposed by Einstein himself \cite{telep4}. Very recently, a reformulation of $f(T)$ gravity was studied by Krssak and Saridakis \cite{krssak}. They used both tetrad and spin connection as dynamical variables and gave a fully covariant, consistent, and frame-independent version of $f(T)$ gravity. Suitable modifications of the theory can be compared with the effective fluid approach to fix the gravitational lagrangian such that a late time accelerated expansion is realized. For more reviews, cosmological significance and recent progress on teleparallel theories we refer the reader to the reviews of Nojiri and Odintsov \cite{Nojiri:2010wj}, Capozziello and De-Laurentis \cite{Capozziello:2011et}, Faraoni and Capozziello \cite{Capozziello:2010zz}, Bamba and Odintsov \cite{Bamba:2015uma}, Nojiri, Odintsov and Oikonomou \cite{Nojiri:2017ncd}, Bamba, Capozziello, Nojiri and Odintsov \cite{Bamba:2012cp}. \medskip

The advantage of modified teleparallel gravity is that it produces field equations that remain second order irrespective of the choice of Lagrangian, a fact which also remains true for a number of extensions of the model. The second order nature means that the gravitational polarization modes remain identical to those in GR, discussed by Farrugia, Said, Gakis and Saridakis \cite{Farrugia:2018gyz}. Therefore the modifications of the theory can be contained within the cosmological regime without interfering with the GR strong field regime of the theory. Now, in the cosmological setup, this theory has received significant attention in recent years with several works on producing accelerating late time solutions, for instance by Bengochea and Ferraro \cite{ben}, Linder \cite{Linder:2010py}. The goal of these works is to replace the cosmological constant component of GR while preserving the matter epoch contribution. It was shown by Wu and Yu\cite{wuyu1} that the Universe can evolve from a radiation dominated era into a matter dominated one, and finally enter an exponential expansion phase under the scope of a power law $f(T)$ gravity. Studies on the observational constraints of modified teleparallel gravity have also been conducted by Wu and Yu\cite{wuyu2} where $WMAP$ data is used to constrain the model parameters. In general, modified teleparallel gravity bodes well against observational data in the cosmological regime as discusssed by Capozziello, Cardone, Farajollahi and Ravanpak \cite{capo1}. Very recently Nunes, Pan and Saridakis \cite{Nunes:2018evm} compared these models against Planck data as well. Stability of extended teleparrallel theory has also been studied in details by Bamba, Geng, Lee and Luo \cite{bamba1}, Behboodi, Akhshabi and Nozari \cite{behboodi}, Paliathanasis, Barrow and Leach \cite{palia}, Paliathanasis, Said and Barrow \cite{Paliathanasis:2017htk}, Farrugia and Said \cite{Farrugia:2016qqe}. On an astrophysical scale, modified teleparallel gravity already has a weak field solution discussed by Ruggiero and Radicella\cite{ruggiero}, which has been used to determine the galactic velocity profile by Finch and Said\cite{Finch:2018gkh}, using a small modification to the $f(T)$ Lagrangian. \medskip

Principally, any modification of gravity must first evade deviations at solar system scales since the GR field equations predict to a very high degree the results of these tests. Despite being an alternative theory of gravity, teleparallel gravity can produce the equivalent field equations as GR for a specific Lagrangian. As with other modifications to gravity, solar system tests allow for a broad range of Lagrangian terms that only take effect cosmologically. In Ref.\cite{said}, these tests were considered in turn within the context of a power-law style modification with a Lagrangian $f(T) \sim T + \alpha T^n$. In this work, it was found that such models do indeed pass all solar system tests for a broad range of model parameters. Indeed, works of this nature can set limits on setting bounds on model parameters. \medskip

An important role in generalized teleparallel theories is played by the choice of the functional form of $f (T)$. The lack of any firmly established theoretical constraint leaves open a way for wide range of possibilities. One can put the constraints on the choice of function by comparing different theoretical predictions with the observational data. Alternatvely one can use a principle of reverse determination where one reconstructs the expected form of the theory from the field equations, starting from a proper cosmological dynamics. Overall in the context of modifications of gravity, reconstruction schemes are studied exhaustively over the years (for a very good overview, we refer to the works of Nojiri and Odintsov \cite{nojireco1, nojireco2, nojireco3}). However, in $f (T)$ theories, the avenue of reconstruction of lagrangian from cosmological parameters is relatively new. Myrzakulov studied reconstruction of $f (T)$ models from a cosmic dynamics giving late time accelerated expansion \cite{myrza1}. Dent, Dutta and Saridakis \cite{dent} gave a general formalism for reconstructing $f (T)$ models for any given dynamical dark energy scenario. A thorough reconstruction analysis of the $f (T)$ models and the conditions for the equivalence of $f(T)$ models with purely kinetic k-essence was studied recently by Myrzakulov \cite{myrza2}. Bamba, Myrzakulov, Nojiri and Odintsov \cite{bamba2} studied reconstruction of an $f (T)$ model, realizing inflation in the early universe, the $\Lambda CDM$ model, Little Rip cosmology and the Pseudo-Rip cosmology. More recent attempts of studying cosmological reconstruction in modified theories of gravity involve using a massive scalar field (Chakrabarti et. al. \cite{scjs, scjskbfR}.  \medskip

The present work deals with the field equations of a covariant formulation of $f (T)$ gravity and attempts to reconstruct the lagrangian from the cosmological jerk parameter $j$. The basic concept of a jerk parameter comes from the second order time evolution of Hubble parameter. Hubble parameter is defined as the fractional rate of the expansion of the universe. First order evolution of the hubble parameter $H$ is defined by the deceleration parameter $q$ which was believed to be constant only until recently. The current observational evidence suggests that the universe shifted into a phase of acceleration from a decelerated phase in the recent past and this straightaway compels one to look at the evolution of $q$. This includes the study of a third order time derivative of the scale factor, defined in terms of the jerk parameter. If the time evolution of the jerk parameter $j$ is known, a third order differential equation for the scale factor can be written from the definition and hence one can find the cosmological evolution in it’s exact form. A similar method can not be performed using deceleration parameter as a tool of reconstruction alone, since the exact evolution of deceleration parameter is not known. Moreover, using a higher (third) order differential equation one allows the spectrum of solutions to be much wider than a second order equations. This also carries motivations from very recent works in literature on parametric reconstruction of the jerk parameter from diverse observational data sets \cite{ankan}. Most of the attempts to build up a dark energy model hover around finding one which in the present epoch resembles a $\Lambda CDM$ model. We demonstrate the methodology of reconstruction for a $\Lambda CDM$ model to begin with, for which one has a constant $j$ as $j = 1$. We also present a few different cases where the jerk parameter is considered a variable, although the degree of nonlinearity of the differential equations for such cases is very high. No functional form of $f (T)$ is assumed at the outset. \\

In Ref.\cite{nojiodisaez}, a similar scheme was used in the context of $f(R)$ gravity where $f(R)$ theory forms generally fourth order field equations. Here, the approach also included early- and late-time considerations with interesting results for cosmological models in $f(R)$ theory. This led to a broader review in Ref.\cite{Nojiri:2010wj} that encapsulated the body of work within the $f(R)$ gravity framework. For instance it was proved that any cosmic expansion history $a \sim a_{0} e^{g(t)}$ can be realized as the solution of some specific reconstructed $f(R)$. As a result the need to use with auxiliary scalar fields could be avoided which is somewhat a subtle motivation of the present work as well. Moreover, in the present approach one does not assume the expansion history at the outset. Rather a quantity entirely kinematic in nature provides for the scale factor depending on the evolution of the aforementioned parameter which therefore defines a novelty of the present work on it's own as compared with $f(R)$ reconstruction schemes. The crucial difference comes in through the realisation that modified teleparallel gravity field equations are second order and so don't suffer from some of the issues that are present in $f(R)$ gravity. For this reason, this work may provide a way forward in forming viable cosmological models within the modified teleparallel gravity context. Reconstruction schemes in the present case are therefore far more simple in nature and open for further allied investigations. The examples presented in both the cases mimic a $\Lambda$CDM cosmic evolution, at least as far as late-time acceleration is concerned. In contract to Ref.\cite{Nojiri:2010wj}, we do not restrict our case only for dust, rather we include a perfect fluid with the equation of state being a parameter for our models which is only possible since the field equations are second order in nature. \\

In the present work we first introduce the teleparallel framework in section $II$, while in section $III$ the Friedmann equation is discussed for a flat cosmology. In section $IV$, this reconstruction treatment is explored for a constant jerk parameter with a functional model of $f (T)$ as a result. In section $V$, this is done for a jerk parameter that has an inverse square dependence on the Hubble parameter, while in section $VI$ a jerk parameter that is slowly varying about its current value is investigated. Section $VII$ is then dedicated to exploring the stability of these models. In section $VIII$ we discuss the phase space portraits for teleparrallel cosmology in brief. Finally, the results are summarized in section $IX$ with a short discussion on further work. Unless stated otherwise, geometric units are used where $G = 1 = c$. Also Latin indices are used to refer to local inertial coordinates while Greek ones are used to refer to global coordinates.

\section{II. Introduction to \texorpdfstring{$f(T)$}{f(T)} Gravity}

Teleparallel gravity carries a fundamenatal disctinction from curvature based descriptions of gravity. Here, the Levi-Civita connection is replaced with the Weitzenbock connection, $\widehat{\Gamma}^{\alpha}_{\mu\nu}$, which is a curvatureless connection and is given by
\begin{equation}\label{eq:weitzenbockdef}
\widehat{\Gamma}^{\rho}_{\nu\mu} := \dut{e}{a}{\rho}\partial_\mu \udt{e}{a}{\nu} + \dut{e}{a}{\rho}\udt{\omega}{a}{b\mu}\udt{e}{b}{\nu}.
\end{equation}
$\udt{e}{a}{\rho}$ is the tetrad field which represents the transformations between arbitrary points in the tangent space. $\dut{e}{a}{\mu}$ represents the tetrad inverse. To account for local Lorentz invariance in the formalism, the spin connection $\udt{\omega}{a}{b\mu}$ is introduced. Together the tetrad and spin connection specify the frame analogous to the metric tensor scenario. The Latin indices represent coordinates on the tangent space and the Greek indices represent general manifold coordinates.\\

The metric is constructed out of the tetrad fields through \cite{caireview}
\begin{equation}
g_{\mu\nu}\left(x\right):= \udt{e}{a}{\mu}\left(x\right)\udt{e}{b}{\nu}\left(x\right)\eta_{ab},
\end{equation}
where $\eta_{ab}$ is the Minkowski metric. In the following work, we suppress the expression of the local position $x$. \\

In $f(T)$ theories, the Riemann tensor of GR  is replaced with the torsion tensor \cite{krssak}
\begin{equation}\label{eq:torsiontensordef}
\udt{T}{a}{\mu\nu} := \partial_\mu \udt{e}{a}{\nu} - \partial_\nu \udt{e}{a}{\mu} + \udt{\omega}{a}{b\mu}\udt{e}{b}{\nu} - \udt{\omega}{a}{b\nu}\udt{e}{b}{\mu},
\end{equation}
which vanishes when there is no gravity. Together with the torsion tensor, we also define the superpotential tensor as
\begin{equation}\label{eq:superpotentialdef}
\dut{S}{a}{\mu\nu} := \frac{1}{2}\left(\udt{K}{\mu\nu}{a}+\dut{e}{a}{\mu}\udt{T}{\alpha\nu}{\alpha}-\dut{e}{a}{\nu}\udt{T}{\alpha\mu}{\alpha}\right).
\end{equation}
The formulation is laid in such a way that the superpotential plays a role closely related to an energy-momentum tensor of gravity \cite{Aldrovandi:2013wha}. Moreover, $\udt{K}{\mu\nu}{a}$ defines the contorsion tensor, written as \cite{caireview}
\begin{equation}\label{eq:contorsiondef}
\udt{K}{\mu\nu}{a} := \dfrac{1}{2} \left(\dut{T}{a}{\mu\nu} + \udt{T}{\nu\mu}{a} - \udt{T}{\mu\nu}{a}\right).
\end{equation}

Together, the torsion tensor in Eq.\eqref{eq:torsiontensordef} and the superpotential tensor in Eq.\eqref{eq:superpotentialdef} combine to produce the so-called teleparallel equivalent of general relativity (TEGR). The theory is exactly equivalent to GR at the level of field equations in the classical regime. The Torsion scalar is defined as
\begin{equation}\label{eq:torsionscalardef}
T := \dut{S}{a}{\mu\nu}\udt{T}{a}{\mu\nu}.
\end{equation}

The TEGR Lagrangian then becomes
\begin{equation}\label{eq:teleparallel-action}
S = \dfrac{1}{16\pi G} \int d^4x \: e \: T + \int d^4x \: e \: \mathcal{L}_m,
\end{equation}
where $e = \det\left(\dut{e}{\mu}{A}\right) = \sqrt{-g}$, and $\mathcal{L}_m$ represents the matter Lagrangian. Now, in a similar manner of the generalization of Ricci scalar in $f(R)$ theories, the teleparrallel action can be generalized into a more general functional form, $f(T)$, to write the gravitational action as
\begin{equation} \label{eq:general-action}
S = \dfrac{1}{16\pi G} \int d^4x \: e \: \left[T + f(T)\right] + \int d^4x \: e \: \mathcal{L}_m.
\end{equation}

Variation of the action with respect to the tetrads, one finds the field equations of the theory as \cite{krssak}
\begin{align}\label{eq:general-field-equations}
&\left(1+f_T\right) \left[e^{-1} \partial_\nu \left(e \dut{S}{a}{\mu\nu}\right) - \udt{T}{b}{\nu a} \dut{S}{b}{\nu\mu} + \udt{\omega}{b}{a\nu}\dut{S}{b}{\nu\mu} \right]\nonumber\\
&+ f_{TT} \dut{S}{a}{\mu\nu} \partial_\nu T + \dut{e}{a}{\mu} \left(\dfrac{f+T}{4}\right) = 4\pi G \dut{e}{a}{\mu} \stackrel{\textbf{em}}{\dut{T}{\mu}{\rho}}.\nonumber\\
\end{align}

The stress-energy tensor in terms of the matter Lagrangian is defined as $\stackrel{\textbf{em}}{\dut{T}{\beta}{\rho}} = \dfrac{1}{e} \udt{e}{a}{\beta} \dfrac{\delta \left(e \mathcal{L}_{\text{m}}\right)}{\delta \udt{e}{a}{\rho}}$.

\section{III. Flat, isotropic and homogeneous universe in \texorpdfstring{$f(T)$}{f(T)} gravity}
We consider a spatially flat, isotropic and homogeneous Friedmann-Lemaitre-Robertson-Walker (FLRW) metric as
\begin{equation}\label{eq:flat-FLRW}
ds^2 = dt^2 - a^2(t)\left(dx^2+dy^2+dz^2\right).
\end{equation}
$a(t)$ is the cosmological scale factor. This can be equivalently written in terms of a diagonal tetrad as
\begin{equation}\label{eq:diag-tetrad}
\dut{e}{\mu}{a} = \text{diag}\left(1,a(t),a(t),a(t)\right),
\end{equation}
where the spin connection turns out to vanish for this choice of the tetrad. It is straightforward to check that for such a tetrad, the torsion scalar $T = -6H^2$, where $H(t) = \frac{\dot{a}}{a}$. From Eq. \eqref{eq:general-field-equations}, one finds the field equations for the current setup as
\begin{align}
& f-T-2Tf_T = 2 \kappa^2 \rho, \label{eq:00-zero} \\ 
& \dot{H} = -\dfrac{\kappa^2 \left(\rho+p\right)}{2\left(1+f_T + 2T f_{TT}\right)}, \label{eq:trace-zero}
\end{align}
where the second equation is written as a combination of the first and second Friedmann equations. \\

The trace of the field equations leads one to the continuity equation
\begin{equation}\label{eq:continuity-zero}
\dot{\rho} + 3H\left(\rho+p\right) = 0,
\end{equation}
which is equivalent to its GR analogue. Defining an equation of state parameter $\omega$ such that
\begin{equation}\label{eq:eos-matter}
p = \omega\rho,
\end{equation}
Eq.\eqref{eq:continuity-zero} can be solved straightaway to write
\begin{equation}\label{eq:continuity-zero-solution}
\rho = \rho_0 {a(t)}^{-3(1 + \omega)}.
\end{equation}

These equations can be interpreted as a TEGR cosmology with an additional cosmic driver, where the exotic components would be defined through
\begin{align}
&\kappa^2 \rho_{\text{exo}} := Tf_T - \dfrac{f}{2}, \label{eq:def-DE-energy-density} \\
&\kappa^2 p_{\text{exo}} := -\kappa^2 \rho_{\text{exo}} + 2\dot{H}\left(f_T+2T f_{TT}\right). \label{eq:def-DE-pressure}
\end{align}
Naturally, these components can be related by an Equation of State, $\omega_{\text{exo}}$ defined as
\begin{align}
&\omega_{\text{exo}} \equiv \dfrac{p_{\text{exo}}}{\rho_{\text{exo}}} = -1 - 4\dot{H}\dfrac{f_T+2T f_{TT}}{f-2Tf_T}.
\end{align}

Using Eq.\eqref{eq:00-zero} and \eqref{eq:trace-zero} again, the EoS parameters can be written as
\begin{align}\label{eq:def-EoS-DE}
\omega_{\text{exo}}&= -1 \nonumber\\
&+ (1 + \omega) \dfrac{\left(f-T-2T f_T\right)\left(f_T+2T f_{TT}\right)}{\left(1 + f_T + 2T f_{TT}\right)\left(f-2Tf_T\right)}.\nonumber\\
&
\end{align}

Using the fact that $T = -6 H^2$, one can easily write Eq.\eqref{eq:def-EoS-DE} as a function of $H$ so that
\begin{equation}\label{eq:def-Eos-DE-H}
\omega_{\text{exo}} = -1 - \frac{(1 + \omega)}{12}\frac{\Big(f + 6H^2 -H f_{H}\Big) f_{HH}}{\Big( 1 - \frac{1}{12}f_{HH}\Big)\Big(f -H f_{H}\Big)}.
\end{equation}

Therefore, the effective Friedmann equations in their usual form alongwith an exotic fluid part coming from the torsion scalar can be written as 
\begin{align}
-T &= 2\kappa^2 \left(\rho + \rho_{\text{exo}}\right), \\ 
2\dot{H} &= -\kappa^2 \left(\rho+p+\rho_{\text{exo}} + p_{\text{exo}}\right).
\end{align}

\section{IV. Reconstruction from a constant jerk}
In this section, we study cosmological reconstruction from a constant jerk parameter. For simplicity, we set this to unity, so that 
\begin{equation}\label{jerkeq1}
j = \frac{d^{3}a(t)}{dt^3} \frac{a(t)^2}{(\frac{da(t)}{dt})^3} = 1.
\end{equation}

The general solution of Eq. \eqref{jerkeq1} can be given by
\begin{equation}\label{scale_constjerk}
a(t) = \Big(A e^{\lambda t} + B e^{-\lambda t}\Big)^{\frac{2}{3}},
\end{equation}
where $A$, $B$ and $\lambda$ are constant parameters that would be fixed by the growth profile of the universe. It is straightforward to note that, the term $A e^{\lambda t}$ dominates over $B e^{-\lambda t}$ at late times. Also, the Hubble parameter can be calculated as

\begin{equation}
H(t) = \frac{2 \lambda}{3} \frac{\Big(A e^{\lambda t} - B e^{-\lambda t}\Big)}{\Big(A e^{\lambda t} + B e^{-\lambda t}\Big)}.
\end{equation}

\begin{figure}
\begin{center}
\includegraphics[width=0.40\textwidth]{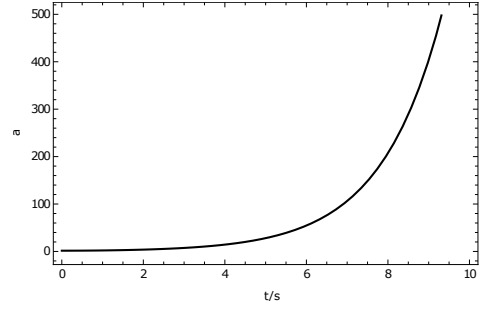}
\caption{Evolution of the scale factor for constant jerk parameter Eq.\eqref{scale_constjerk} with cosmic time, for $\lambda = 1$, $A = 0.2$ and $B = -0.1$.}
\label{a_constant_jerk}
\end{center}
\end{figure}

\begin{figure}
\begin{center}
\includegraphics[width=0.40\textwidth]{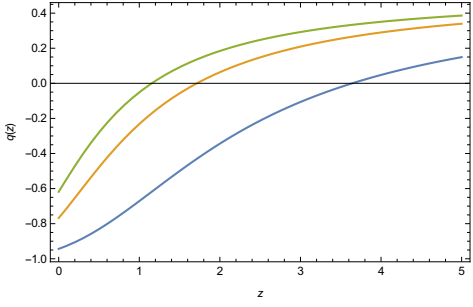}
\caption{Evolution of the deceleration parameter for constant jerk parameter Eq. \eqref{scale_constjerk} with redshift $z$, for $\lambda = 1$ and $A = 0.2$. The blue curve represents a choice of the parameter $B = -0.1$ while yellow corresponds to $B = -0.5$ and green refers to $B = -1$.}
\label{q_constant_jerk}
\end{center}
\end{figure}

We plot the time evolution of the scale factor in Fig. (\ref{a_constant_jerk}) to show the accelerating behavior for a particular case, $\lambda = 1, A = 0.2$ and $B = - 0.1$ which generically shows late-time accelerating expansion. Moreover, in Fig. (\ref{q_constant_jerk}), we plot the evolution of the deceleration parameter $q$ as a function of redshift $z$ for the constant jerk case, choosing three different set of parameters, (i) $\lambda = 1$, $A = 0.2$ and $B = -0.1$; (ii) $\lambda = 1$, $A = 0.2$ and $B = -0.5$; (iii) $\lambda = 1$, $A = 0.2$ and $B = -1$. While the particular values of negative deceleration at current times is different, the behavior easily gives an accelerating Universe at late-times. \medskip

\noindent The torsion scalar $T$ can also straightforwardly be determined as
\begin{equation}\label{torsion_sca_def}
T = -6H^2 = -\frac{8\lambda^2}{3}\frac{(Ae^{\lambda t} - Be^{-\lambda t})^2}{(Ae^{\lambda t} + Be^{-\lambda t})^2}.
\end{equation}

For $p = \omega \rho$, Eq.(\ref{eq:00-zero}) and Eq.(\ref{eq:trace-zero}) can be rewritten as
\begin{align}
& f-T-2Tf_T = 2 \kappa^2 \rho, \label{eq:00-zero1} \\ 
& 4\dot{H} \left(1+f_T + 2T f_{TT}\right) = - 2 \kappa^2 (1 + \omega) \rho, \label{eq:trace-zero1}
\end{align}
which can then be combined to get
\begin{align}\label{eqM}
f -& T - 2Tf_T \nonumber\\
&+ \frac{1}{(1+\omega)} [4\dot{H} \left(1+f_T + 2T f_{TT}\right)] = 0.
\end{align}

To proceed, Eq.(\ref{eqM}) must be solved for an $f(T)$ model or, equivalently, for $f(H)$  by using the relation in Eq.(\ref{torsion_sca_def}) between the Hubble parameter and Torsion scalar that will then determine the reconstruction scheme. The terms of this equation can be written in terms of $H$ as
\begin{align}
& \dot{H} = \frac{2\lambda^2}{3} - \frac{3H^2}{2}, \\
& f_T = -\frac{1}{12H} \frac{df}{dH}, \\
& f_TT = \frac{1}{144H^2}\frac{d^2 f(H)}{dH^2} - \frac{1}{144H^3}\frac{df(H)}{dH}.
\end{align} 

Thus, the combined Friedmann equation in Eq.(\ref{eqM}) transforms into
\begin{align}\label{eqM1}
 \frac{d^{2}f(H)}{dH^2} &- \frac{2(1+\omega)H\frac{df(H)}{dH}}{(\frac{4\lambda^2}{9}-H^2)} - \frac{2(1+\omega)f(H)}{(\frac{4\lambda^2}{9}-H^2)} \nonumber\\
&- \frac{2(18 \omega H^2 + 8 \lambda^{2})}{3(\frac{4\lambda^2}{9} - H^{2})} = 0.
\end{align}

In its current form, Eq.(\ref{eqM1}) is intractable, and so we transform it under
\begin{align}
H &= \frac{2\lambda}{3} \cos G, \\
\frac{df(H)}{dH} &= -\frac{3}{2\lambda} \text{cosec} G \frac{df(G)}{dG}, \\
\frac{d^{2}f(H)}{dH^2} &= \frac{9}{4\lambda^2} \text{cosec}^2 G \Bigg[\frac{d^{2}f(G)}{dG^2} - \cot G \frac{df(G)}{dG} \Bigg],
\end{align}
and rewrite Eq. (\ref{eqM1}) as
\begin{align}\label{eqM2}
& \frac{d^{2} f(G)}{dG^2} - \cot G \frac{df(G)}{dG} - 2 (1+\omega) \cot G \frac{df(G)}{dG} \nonumber\\
&- 2 (1+\omega) f(G) - \frac{16}{3} \omega \lambda^2 \cos^{2} G - \frac{16}{3} \lambda^2 = 0. \nonumber\\
&
\end{align}

We study the equation for specific choices of the equation of state parameter $\omega$.
\begin{enumerate}
\item {For $\omega = 0$, i.e., for a dust solution, one can solve Eq.(\ref{eqM2}) to find
\begin{align}
f(H) &= -\frac{8\lambda^2}{3} \nonumber\\
&+ C_{1}\Bigg(1-\frac{3H}{2\lambda}\Bigg)^{\frac{3}{2}} \Bigg(\frac{3H}{2\lambda}-1\Bigg)^{\frac{1}{2}} \nonumber\\
&+ C_{2}\frac{3H}{2\lambda}\Bigg(1-\frac{3H}{2\lambda}\Bigg)^{\frac{3}{2}} \Bigg(\frac{3H}{2\lambda}-1\Bigg)^{\frac{1}{2}}.
\end{align}
or in terms of $H = (-\frac{T}{6})^{\frac{1}{2}}$,
\begin{align}
f(T) &= -\frac{8\lambda^2}{3} \nonumber\\
&+ C_{1}\Bigg(1-\frac{3(-\frac{T}{6})^{\frac{1}{2}}}{2\lambda}\Bigg)^{\frac{3}{2}} \Bigg(\frac{3(-\frac{T}{6})^{\frac{1}{2}}}{2\lambda}-1\Bigg)^{\frac{1}{2}} \nonumber\\
&+ C_{2}\frac{3(-\frac{T}{6})^{\frac{1}{2}}}{2\lambda}\Bigg(1-\frac{3(-\frac{T}{6})^{\frac{1}{2}}}{2\lambda}\Bigg)^{\frac{3}{2}} \nonumber\\
&\Bigg(\frac{3(-\frac{T}{6})^{\frac{1}{2}}}{2\lambda}-1\Bigg)^{\frac{1}{2}}.
\end{align}
}
\item {For $\omega = -1$, meaning the EoS of dark energy \cite{Aghanim:2018eyx}, the reconstruction procedure gives 
\begin{equation}
f(H) = C_{1} + \frac{3C_{1}}{2\lambda}H + 6H^2. 
\end{equation}
or
\begin{equation}
f(T) = C_{1} + \frac{3C_{1}}{2\lambda} \Big(-\frac{T}{6}\Big)^{\frac{1}{2}} - T. 
\end{equation}
}
\item {For $\omega  \gtrsim -1$, one can find out that
\begin{align}\nonumber
f(H) &= C_{1} + \frac{3C_{2}}{2\lambda}H + \frac{1}{30}\Bigg[77\lambda^2 + 77\lambda^2 \Bigg(\frac{3H}{2\lambda}\Bigg)^2 \nonumber\\
&-3\lambda^{2} \Bigg(\frac{3H}{2\lambda}\Bigg) \ln(1-\frac{3H}{2\lambda}) + 3\lambda^{2}\Bigg(\frac{3H}{2\lambda}\Bigg) \nonumber\\
&\ln\left(1+\frac{3H}{2\lambda}\right) + 3\lambda^{2} \ln\Bigg(\frac{9H^2}{4\lambda^2} - 1\Bigg)\Bigg].
\end{align}
or
\begin{align}\nonumber
f(T) &= C_{1} + \frac{3C_{2}}{2\lambda}\Big(-\frac{T}{6}\Big)^{\frac{1}{2}} + \frac{1}{30}\Bigg[77\lambda^2 \nonumber\\
&+ 77\lambda^2 \Bigg(\frac{3\Big(-\frac{T}{6}\Big)^{\frac{1}{2}}}{2\lambda}\Bigg)^2 
-3\lambda^{2} \Bigg(\frac{3\Big(-\frac{T}{6}\Big)^{\frac{1}{2}}}{2\lambda}\Bigg) \nonumber\\
&\ln\Bigg(1 -\frac{3\Big(-\frac{T}{6}\Big)^{\frac{1}{2}}}{2\lambda}\Bigg) + 3\lambda^{2}\Bigg(\frac{3\Big(-\frac{T}{6}\Big)^{\frac{1}{2}}}{2\lambda}\Bigg) \nonumber\\
&\ln\Bigg(1 +\frac{3\Big(-\frac{T}{6}\Big)^{\frac{1}{2}}}{2\lambda}\Bigg) + 3\lambda^{2} \ln\Bigg(\frac{9\Big(-\frac{T}{6}\Big)}{4\lambda^2} \nonumber\\
&- 1\Bigg)\Bigg].
\end{align}
}
\end{enumerate}

\begin{figure}
\begin{center}
\includegraphics[width=0.40\textwidth]{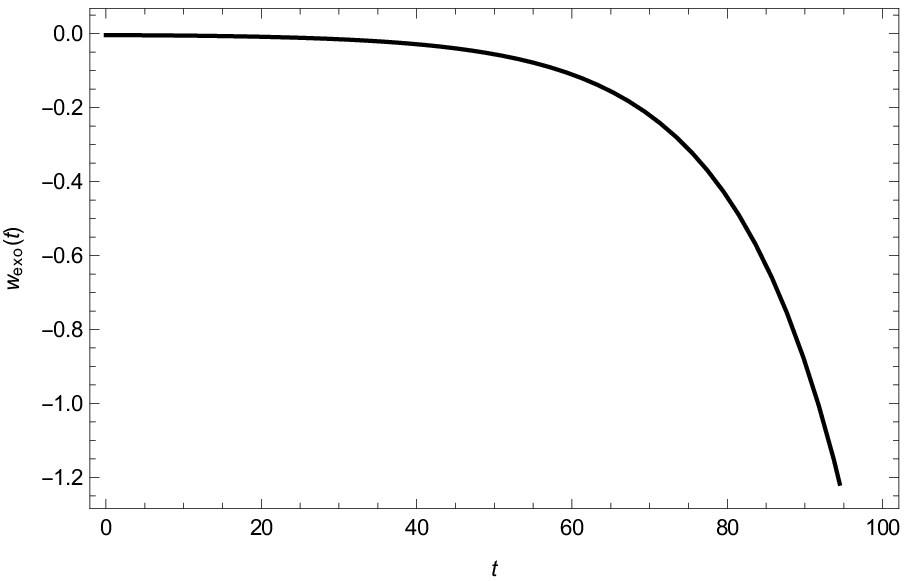}
\includegraphics[width=0.40\textwidth]{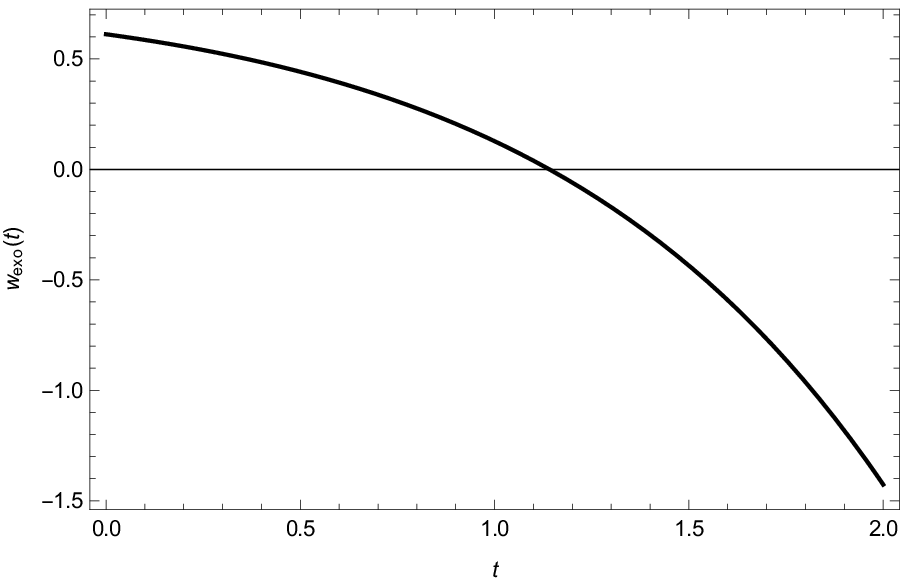}
\includegraphics[width=0.40\textwidth]{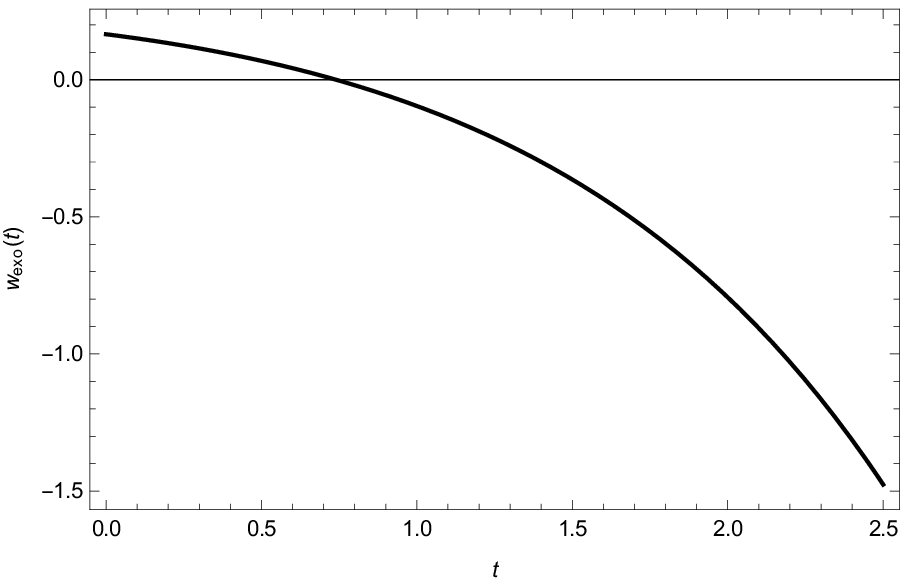}
\includegraphics[width=0.40\textwidth]{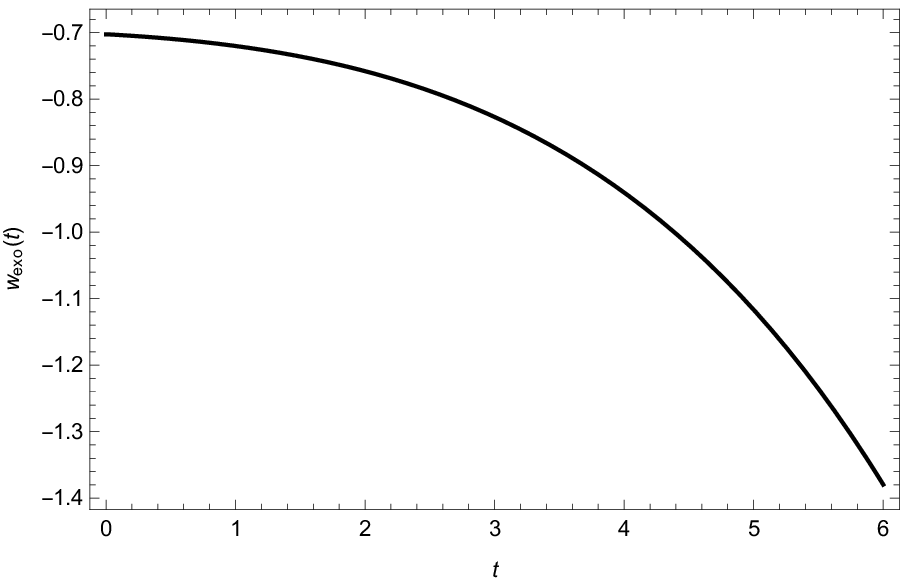}
\caption{Equation of state of the exotic fluid for reconstructed $f(T)$ model from a constant jerk parameter. $\lambda$ is fixed at $0.035$ and $A = 2$, $B = 1$. The top graph represents plot of $\omega = 0$, the second from the top is for $\omega = 1$, third is for $\omega = \frac{1}{2}$, while the graph on the bottom is for $\omega = -\frac{1}{2}$.}
\label{EOS_exotic_constant_jerk}
\end{center}
\end{figure}

However, we also make an attempt to comment on the general behavior of $f(T)$ (or $f(H)$) without any \textit{a priori} choice of the EoS but making certain reasonable approximations at late times. We assume that the parameter $\lambda$ is small such that the terms involving $\lambda^2$ in Eq.(\ref{eqM2}) can be neglected with respect to the other terms. Thus the equation for reconstruction is slightly simplified, and a solution for $f(H)$ can be written in terms of a hypergeometric function such that
\begin{align}\label{generalfH}
f(H) &= \frac{3C_{1}}{2\lambda}H - \frac{3C_{2}}{2\lambda}H \Bigg[\frac{3H}{2\lambda} {_2}F_{1} \Big(\frac{1}{2}, -\omega, \nonumber\\
&\frac{3}{2}, \frac{9H^2}{4\lambda^2}\Big) + \frac{2\lambda}{3H} {_2}F_{1} \Big(-\frac{1}{2}, -\omega, \frac{1}{2}, \nonumber\\
&\frac{9H^2}{4\lambda^2}\Big) \Bigg].
\end{align} 

In terms of $H = (-\frac{T}{6})^{\frac{1}{2}}$,

\begin{align}
f(T) &= \frac{3C_{1}}{2\lambda}\Big(-\frac{T}{6}\Big)^{\frac{1}{2}} - \frac{3C_{2}}{2\lambda}\Big(-\frac{T}{6}\Big)^{\frac{1}{2}} \nonumber\\
&\Bigg[\frac{3(-\frac{T}{6})^{\frac{1}{2}}}{2\lambda} {_2}F_{1} \Big(\frac{1}{2}, -\omega, \frac{3}{2}, \frac{9(-\frac{T}{6})}{4\lambda^2}\Big) \nonumber\\
&+ \frac{2\lambda}{3(-\frac{T}{6})^{\frac{1}{2}}} {_2}F_{1} \Big(-\frac{1}{2}, -\omega, \frac{1}{2}, \frac{9(-\frac{T}{6})}{4\lambda^2}\Big) \Bigg].
\end{align}

As discussed in \textsection. III, the EoS of the exotic fluid can be written as a function of torsion scalar and therefore as a function of $H$ as given by Eq. (\ref{eq:def-Eos-DE-H}). We plot the exotic equation of state as a function of cosmic time in Fig. (\ref{EOS_exotic_constant_jerk}). The plots vary over matter EoS parameters. It also scales for different choices of $\lambda$, however, the qualitative behavior remains the same. For all the cases, $\omega_{exo}$ at early time lies within the range $-1 < \omega_{exo} < 1$. However, at late times, the equation of state approaches a phantom behavior for this analysis. This is interesting given the suggestions by the recent Planck collaboration in this regard \cite{Aghanim:2018eyx}. \\

\section{V. Reconstruction from a variable jerk: Inverse Hubble Parameter}
In this section we study the cosmological dynamics for a variable jerk such that the jerk parameter is proportional to the Hubble parameter by an inverse square relation. In general equation, the jerk parameter can be represented by
\begin{equation}
j = \frac{d^{3}a(t)}{dt^3} \frac{a(t)^2}{(\frac{da(t)}{dt})^3} = g(t),
\end{equation}
which is extremely non-trivial to tackle without any \textit{a priori} assumptions on the function $g(t)$. We assume $g(t)$ to be a function of the Hubble parameter $H$. A particular case in this regard is where the jerk parameter is proportional to the inverse square of this parameter, meaning that
\begin{equation}\label{jerkeq3}
\frac{d^{3}a(t)}{dt^3} \frac{a(t)^2}{(\frac{da(t)}{dt})^3} = \frac{s^{2}}{H^2},
\end{equation}
where $s$ is an arbitrary constant. \medskip

Solving equation Eq.(\ref{jerkeq3}), the solution turns out to be
\begin{equation}\label{scale3}
a(t) = \frac{1}{s}\Big(m e^{s t} - n e^{-s t}\Big) + p.
\end{equation}
$m$, $n$, $s$ and $p$ are constant parameters. 

\begin{figure}
\begin{center}
\includegraphics[width=0.40\textwidth]{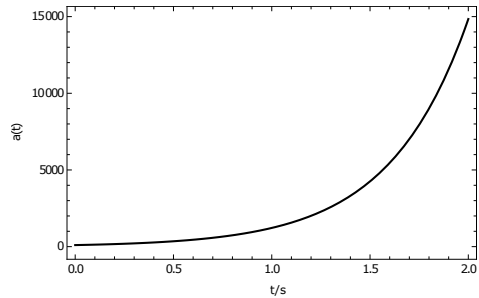}
\caption{Scale factor as a function of cosmic time for jerk parameter in Eq.(\ref{jerkeq3}); $m >> n$, $s = 1$, $p = 0.5$.}
\label{varjerk2_scale_fac}
\end{center}
\end{figure}

In Fig.(\ref{varjerk2_scale_fac}), we plot the scale factor as a function of cosmic time to visualize the late-time cosmic acceleration, for a particular set of representative parameters. For example, if $s$ is fixed to unity and $p$ is chosen to be $0.5$, for a choice of $m >> n$ we expect the exponential term to dominate at late times.

\subsection{Comparison with Supernovae Data}
In section $IV$, we reconstructed cosmological models from a constant value of jerk $j = 1$, which gives exacty the $\Lambda$CDM cosmic history, with already a known plethora of articles on the comparison of model parameters with observational data. However, the present section has dealt with a new ansatz on the aforementioned jerk parameter as a function of hubble parameter. Thus, we fit the available Supernova data $Union2.1$ of Hubble free luminosity distance with the results obtained from the model given by Eq. (\ref{scale3}). \\

At the present time, $a_{0} = 1$ and $\dot{a}_{0}= H_{0}$ which is the present day value of the Hubble parameter. Therefore from Eq. \ref{scale3}, we have
\begin{equation}
H_{0}^{2} = s(1-p)^{2} + 4mn.
\end{equation}
Defining a dimensionless parameter $\mu = \frac{s(1 - p)^{2}}{H_{0}^{2}}$ and writing $\frac{4mn}{H_{0}^{2}} = 1 - \mu$, we rewrite the hubble parameter as 

\begin{equation}\label{hubblemodel2}
H(a)= \frac{H_0}{a} \Bigg[ (1 - \mu) + \mu\frac{(a - p)^{2}}{p^{2}}\Bigg]^{\frac{1}{2}}.
\end{equation}

We define the {\it Hubble free luminosity distance} $d_L$ using $D_L=c H_0^{-1} d_L$, where $D_L$ is the luminosity distance, $H_0$ is the present day observed value of the Hubble parameter and $c$ is the speed of light. The expression of the Hubble free luminosity distance is given by

\begin{equation}
d_L(z)=(1+z)\int^{z}_{0}dz' \frac{H_0}{H(z')} .
\end{equation}    

Using Eq.(\ref{hubblemodel2}) this can be written as 
\begin{eqnarray}\nonumber \label{lumidis}
&& d_L(z;\mu) = \\&&
 (1+z) \int^z_0 \frac{\frac{dz'}{(1+z')}}{\Bigg[(1-\mu)+\frac{\mu}{p^{2}}\Bigg(\frac{1}{(1+z')} - p\Bigg)^{2}\Bigg]^{\frac{1}{2}}}.
\end{eqnarray}

We fit the available Supernova data $Union2.1$ using the model Eq.(\ref{lumidis}) and obtain the best fit value of the parameter $\mu \sim 0.39$ for $p = \frac{1}{2}$. For the analyzed data, we impose a constraint such that $\chi^2_{min}/d.o.f = \chi^2_{min}/(N-n)\lesssim 1$ ($N$: number of data points, $n$: number of parameters) such that the fitting is good and the data are consistent with the considered model. For the present model $\chi^2_{min}/d.o.f \sim 0.93$.

\begin{figure}
\begin{center}
\includegraphics[width=0.40\textwidth]{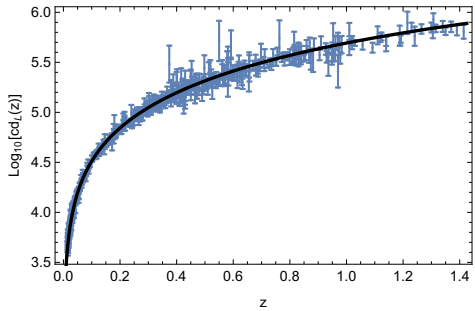}
\caption{Observed $SNe Ia$ Hubble free luminosity distance data along with the fitted curve for the best fit parameter value $\mu \sim 0.39$. $\chi^2_{min}/d.o.f = \chi^2_{min}/(N-n) \sim 0.93$ and therefore the fitting is good.}
\label{Supernovae}
\end{center}
\end{figure}

\begin{figure}
\begin{center}
\includegraphics[width=0.40\textwidth]{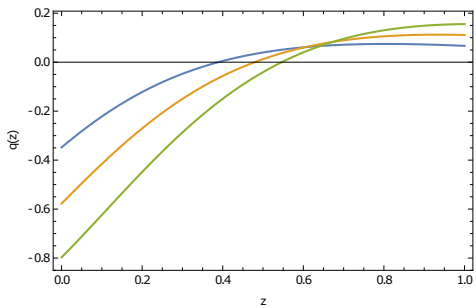}
\caption{Deceleration parameter, $q$, as a function of redshift for jerk parameter in Eq.(\ref{jerkeq3}); Blue is $s = 0.8$, Yellow is $s = 1.0$ and Green is $s = 1.2$. $m$ and $n$ take the same values as in Fig.(\ref{varjerk2_scale_fac})}
\label{varjerk2_q}
\end{center}
\end{figure}

In Fig.(\ref{varjerk2_q}), we plot the evolution of the deceleration parameter $q$ as a function of redshift $z$ for three different choices of parameters. $m$ and $n$ are similarly chosen as in Fig.(\ref{varjerk2_scale_fac}), while the parameter $s$ is varied with choices $0.8$, $1.0$ and $1.2$. As expected, for these values we regain generic deceleration at current times. \medskip

\begin{figure}
\begin{center}
\includegraphics[width=0.40\textwidth]{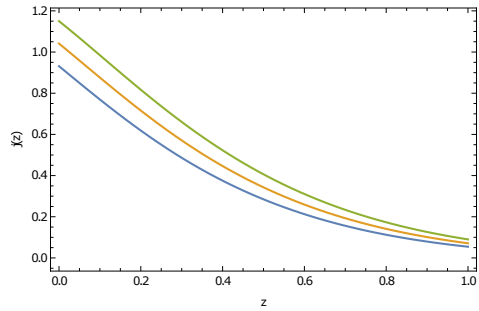}
\caption{Jerk parameter, $j$, as a function of redshift for jerk parameter in Eq.(\ref{jerkeq3}); Blue is $s = 0.85$, Yellow is $s = 0.90$ and Green is $s = 0.95$.}
\label{varjerk2_j}
\end{center}
\end{figure}

In Fig.(\ref{varjerk2_j}), we plot the evolution of the jerk parameter $j$ as a function of redshift $z$ for these three different choices of parameters. $m$ and $n$ are similarly chosen as in Fig.(\ref{varjerk2_scale_fac}), while the parameter $s$ is varied with choices $0.85$, $0.90$ and $0.95$.  These plots are very interesting since they all return jerk parameters at current times close to unity which is not very dissimilar to its observational value. \medskip

Together, Fig.(\ref{varjerk2_q}) and Fig.(\ref{varjerk2_j}) can produce current observational values for specific choice of parameters such that $n << m$. In this limit, the Hubble parameter can be written approximately as
\begin{equation}
\dot{H} = s H - H^2.
\end{equation}

This approximation is then used in the combined Friedmann relation in Eq.(\ref{eqM}), but the approximation turn out to give
\begin{align}
& \dot{H} = s H - H^2, \\
& f_T = -\frac{1}{12H} \frac{df}{dH}, \\
& f_TT = \frac{1}{144H^2}\frac{d^2 f(H)}{dH^2} - \frac{1}{144H^3}\frac{df(H)}{dH}.
\end{align}

\noindent Thus Eq.(\ref{eqM}) in this setup becomes
\begin{align}\label{eqM4}
\frac{d^{2}f(H)}{dH^2} &+ \frac{3(1+\omega)}{(s - H)}\frac{df(H)}{dH} - \frac{3(1+\omega)f(H)}{(s - H)H}
\nonumber\\
&- \frac{18(1+\omega)H}{(s-H)} -12 = 0.
\end{align}

As in the previous scenario, to analytically solve Eq.(\ref{eqM4}), we implement the transformation
\begin{align}
H &= s \cos^2 (G), \\
(s - H) &= s \sin^2 (G), \\
\frac{df(H)}{dH} &= -\frac{1}{s} \text{cosec} (2G) \frac{df(G)}{dG}, \\
\frac{d^2 f(H)}{dH^2} &= \frac{1}{s^2} \Big( \text{cosec}^2 (2G) \frac{d^2 f(G)}{dG^2} \nonumber\\
&- 2 \text{cosec}^2 (2G) cot (2G) \frac{df(G)}{dG}\Big),
\end{align}
and rewrite Eq. \eqref{eqM4} as

\begin{align}\label{eqM5a}
\frac{d^{2} f(G)}{dG^2} &- (2 \cot (2G) + 6(1 + \omega)\cot(G)) \frac{df(G)}{dG} \nonumber\\
&- 12 (1+\omega) f(G) - 72 (1 + \omega) s^2 \cos^{4} (G) \nonumber\\
&- 12 s^2 sin^2 (2G) = 0.
\end{align}

Due to the high non-linearity of Eq.(\ref{eqM5a}), a general solution in closed form cannot be found. We solve the equation for different choices of $\omega$ below.
\begin{enumerate}
\item {For $\omega = 0$,
\begin{align}
f(H) &= -\frac{C_2}{2} + \Big(\frac{C_1}{s} - 9s\Big)H \nonumber\\
&+ \Big(6 + \frac{3C_2}{s^2}\Big)H^2 - \frac{C_2}{4s^2}H^3 \nonumber\\
&- \frac{3C_2}{2s} H \ln\Big(\frac{H}{s}\Big).
\end{align}
or in terms of $H = (-\frac{T}{6})^{\frac{1}{2}}$,
\begin{align}
f(T) &= -\frac{C_2}{2} + \Big(\frac{C_1}{s} - 9s\Big)\Big(-\frac{T}{6}\Big)^{\frac{1}{2}} \nonumber\\
&+ \Big(6 + \frac{3C_2}{s^2}\Big)\Big(-\frac{T}{6}\Big) - \frac{C_2}{4s^2}\Big(-\frac{T}{6}\Big)^{\frac{3}{2}} \nonumber\\
&- \frac{3C_2}{2s} \Big(-\frac{T}{6}\Big)^{\frac{1}{2}} \ln\Bigg[\frac{\Big(-\frac{T}{6}\Big)^{\frac{1}{2}}}{s}\Bigg].
\end{align}
}
\item {For $\omega = -1$,
\begin{equation}
f(H) = C_1 + \frac{C_2}{2s}H + 6H^2.
\end{equation}
or
\begin{equation}
f(T) = C_1 + \frac{C_2}{2s}\Big(-\frac{T}{6}\Big)^{\frac{1}{2}} - T.
\end{equation}
}
\item {For $\omega = -0.5$,
\begin{align}
f(H) &= \Big(\frac{C_1}{s} + 18s\Big)H + 6H^2 \nonumber\\
&+ \frac{C_2}{2}\Bigg[ -\Big(1 - \frac{H}{s}\Big)^{1/2} \Big(1 + \frac{2 H}{s}\Big)\nonumber\\
&- \frac{3H}{s} \ln \Bigg(\frac{1 + \Big(1 - \frac{H}{s}\Big)^{1/2}}{\Big(\frac{H}{s}\Big)^{1/2}}\Bigg) \Bigg].
\end{align}
or
\begin{align}
f(T) &= \Big(\frac{C_1}{s} + 18s\Big)\Big(-\frac{T}{6}\Big)^{\frac{1}{2}} - T + \frac{C_2}{2}\nonumber\\
&\Bigg[ -\Big(1 - \frac{\Big(-\frac{T}{6}\Big)^{\frac{1}{2}}}{s}\Big)^{\frac{1}{2}} \Big(1 + \frac{2 \Big(-\frac{T}{6}\Big)^{\frac{1}{2}}}{s}\Big)\nonumber\\
&- \frac{3\Big(-\frac{T}{6}\Big)^{\frac{1}{2}}}{s} \ln \Bigg(\frac{1 + \Big(1 - \frac{\Big(-\frac{T}{6}\Big)^{\frac{1}{2}}}{s}\Big)^{\frac{1}{2}}}{\Big(\frac{\Big(-\frac{T}{6}\Big)^{\frac{1}{4}}}{s^{\frac{1}{2}}}\Big)}\Bigg) \Bigg].
\end{align}
}
\item {For $\omega = 0.5$,
\begin{align}
f(H) &= \Big(\frac{C_1}{s} - \frac{108s}{35}\Big)H + 6H^2 \nonumber\\
&+ \frac{C_2}{70}\Bigg[ -\Big(1 - \frac{H}{s}\Big)^{1/2} \Big(-35 - \frac{388 H}{s} \nonumber\\
&+ \frac{156 H^2}{s^2} - \frac{58 H^3}{s^3} + \frac{10 H^4}{s^4}\Big) \nonumber\\
&- \frac{315 H}{s} \ln \Bigg(\frac{1 + \Big(1 - \frac{H}{s}\Big)^{1/2}}{\Big(\frac{H}{s}\Big)^{1/2}}\Bigg) \Bigg].
\end{align}
or
\begin{align}
f(T) &= \Big(\frac{C_1}{s} - \frac{108s}{35}\Big)\Big(-\frac{T}{6}\Big)^{\frac{1}{2}} - T + \frac{C_2}{70}\nonumber\\
&\Bigg[ -\Big(1 - \frac{\Big(-\frac{T}{6}\Big)^{\frac{1}{2}}}{s}\Big)^{\frac{1}{2}} \Big(-35 - \frac{388 \Big(-\frac{T}{6}\Big)^{\frac{1}{2}}}{s} \nonumber\\
&+ \frac{156 \Big(-\frac{T}{6}\Big)}{s^2} - \frac{58 \Big(-\frac{T}{6}\Big)^{\frac{3}{2}}}{s^3} + \frac{10 \Big(-\frac{T}{6}\Big)^2}{s^4}\Big) \nonumber\\
&- \frac{315 \Big(-\frac{T}{6}\Big)^{\frac{1}{2}}}{s} \ln \Bigg(\frac{1 + \Big(1 - \frac{\Big(-\frac{T}{6}\Big)^{\frac{1}{2}}}{s}\Big)^{\frac{1}{2}}}{\Big(\frac{\Big(-\frac{T}{6}\Big)^{\frac{1}{4}}}{s^{\frac{1}{2}}}\Big)}\Bigg) \Bigg].
\end{align}
}
\item {For $\omega = 1$,
\begin{align}
f(H) &= \Big(\frac{C_1}{s} - \frac{9s}{5}\Big)H + 6H^2 \nonumber\\
&+ \frac{C_2}{20}\Bigg[-10 + \frac{150}{s^2} H^2 - \frac{100}{s^3} H^3 \nonumber\\
&+ \frac{50}{s^4} H^4 -\frac{15}{s^5}H^5 + \frac{2}{s^6} H^6 \nonumber\\
&- \frac{60}{s} H \ln\Big(\frac{H}{s}\Big)\Bigg].
\end{align}
or
\begin{align}
f(T) &= \Big(\frac{C_1}{s} - \frac{9s}{5}\Big)\Big(-\frac{T}{6}\Big)^{\frac{1}{2}} - T \nonumber\\
&+ \frac{C_2}{20}\Bigg[-10 + \frac{150}{s^2} \Big(-\frac{T}{6}\Big) - \frac{100}{s^3} \Big(-\frac{T}{6}\Big)^{\frac{3}{2}} \nonumber\\
&+ \frac{50}{s^4} \Big(-\frac{T}{6}\Big)^2 -\frac{15}{s^5}\Big(-\frac{T}{6}\Big)^{\frac{5}{2}} + \frac{2}{s^6} \Big(-\frac{T}{6}\Big)^3 \nonumber\\
&- \frac{60}{s} \Big(-\frac{T}{6}\Big)^{\frac{1}{2}} \ln\Big(\frac{\Big(-\frac{T}{6}\Big)^{\frac{1}{2}}}{s}\Big)\Bigg].
\end{align}
}
\end{enumerate}

\begin{figure}
\begin{center}
\includegraphics[width=0.40\textwidth]{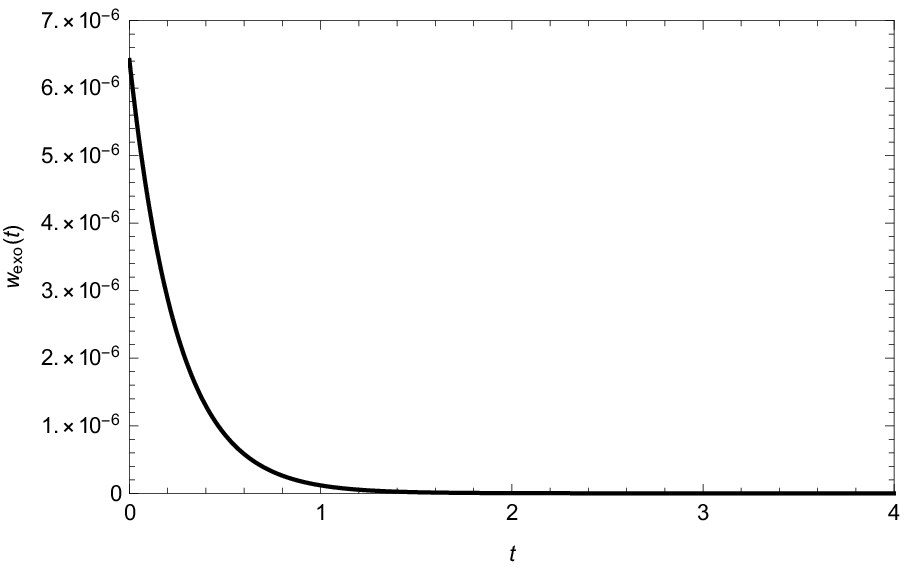}
\includegraphics[width=0.40\textwidth]{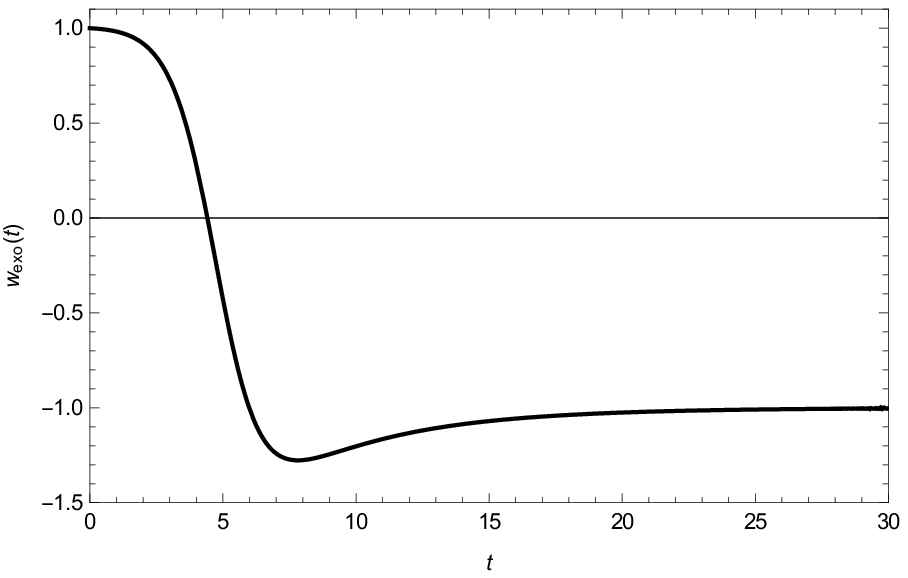}
\includegraphics[width=0.40\textwidth]{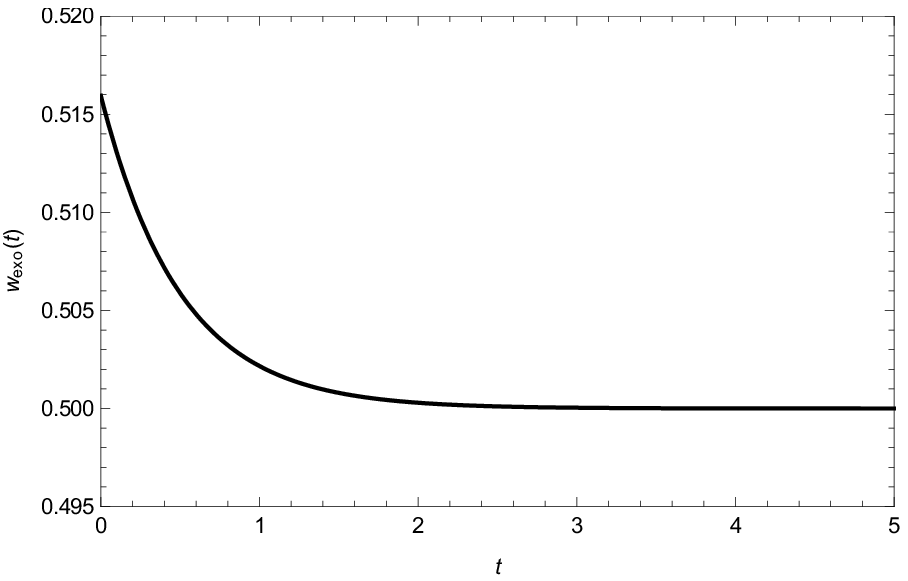}
\includegraphics[width=0.40\textwidth]{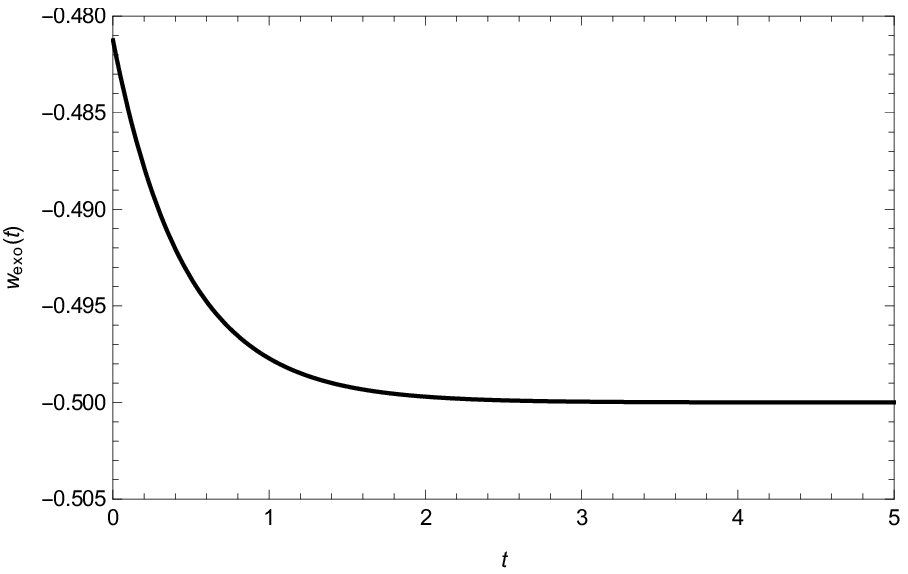}
\caption{EoS of the exotic fluid as a function of cosmic time. $m$ and $n$ are chosen as in Fig.(\ref{varjerk2_scale_fac}), while the parameter $s$ is fixed at $0.1$. The graph on the top shows evolution of $\omega_{exo}$ for $\omega = 0$; the graph second from the top shows the evolution for $\omega = 1$; the graph second from the top shows the evolution for $\omega = \frac{1}{2}$. The graph on the bottom is for $\omega = -\frac{1}{2}$.}
\label{EOS_exotic_variable_jerk_I}
\end{center}
\end{figure}

The exotic EoS, $\omega_{exo}$, given by Eq.(\ref{eq:def-Eos-DE-H}) is plotted as a function of cosmic time in Fig.(\ref{EOS_exotic_variable_jerk_I}). $\omega_{exo}$ depends on the choice of the EoS of the matter fluid $\omega$. It scales for different choices of $m$, $n$ and $s$, however the qualitative behavior remains the same. The top plot shows the evolution of $\omega_{eos}$ for $\omega = 0$. One can see from the graph that the exotic fluid approaches a zero equation of state at late times, although in early times $\omega_{exo} > 0$. The plot second from the top shows the evolution for $\omega = 1$ and as it appears, the evolution starts from $\omega_{eos} = 1$, decreases with time before crossing the $\omega = 0$ mark and thereafter asymptotically settles at $\omega = -1$. As the plot second from the bottom shows, for $\omega = \frac{1}{2}$ the exotic fluid reaches a $\omega_{exo} = \frac{1}{2}$ at late times. Similar behavior is seen for $\omega = -\frac{1}{2}$, shown in the bottom plot, where the $\omega_{exotic}$ asymptotically reaches $-\frac{1}{2}$.

\section{VI. Reconstruction from a variable jerk : A Slowly Varying Jerk Parameter}
In this section, we study a scenario where the jerk parameter is a slowly varying function of redshift $z$.  We represent this scenario by the jerk parameter
\begin{equation}
j(z) = 1 - \epsilon f(z),
\end{equation}
where $\epsilon$ is a small parameter and $f(z)$ is a slowly varying function of $z$. Then, writing the derivative of $f(z)$ as 
\begin{equation}
\frac{df}{dz} = \frac{df}{dH} \frac{dH}{dt} \frac{dt}{d a} \frac{da}{dz},
\end{equation}
and using the relation $z = \frac{1}{a} - 1$, we write the functional form of $f(z)$ as a function of $a$ through the approximation below,
\begin{equation}
f(a) = \epsilon_{1} + \frac{\epsilon_{0}}{a}.
\end{equation}
Here, $\epsilon_1$ is a constant of integration and we have used $\epsilon_{0} = \frac{df}{dz}$, which we take as a very small parameter, such that $f(z)$ is a slowly varying function of $z$. Thus,  it follows that
\begin{equation}\label{eqtripledot}
\frac{\dddot{a}}{\dot{a}}a^3 + (\epsilon \epsilon_1 - 1)\dot{a}^{2}a + \epsilon \epsilon_{0}\dot{a}^2 = 0.
\end{equation} 

After simplification, and two integrations, Eq.\eqref{eqtripledot} turns out to give
\begin{equation}\label{Hequation}
H^2 = \frac{\dot{a}^2}{a^2} = C_{1} a^{\frac{1}{2} - \beta_{2} - 2} + C_{2} a^{\frac{1}{2} + \beta_{2} - 2},
\end{equation}  
where $\beta_{2} = \frac{1}{2}(1 - 8\alpha)^{\frac{1}{2}}$ and $\alpha = (\epsilon\epsilon_{1} - 1)$.  \medskip

This can straightforwardly be integrated to yield a general expression for the scale factor, however,  we are also interested in the expression of $\dot{H}$ in terms of $H$. For such a purpose we use a particular value of $\epsilon_{1}$ for which $\beta_{2} = 0$. From that we can write
\begin{equation}
\dot{H} \sim \lambda H^2,
\end{equation}
where $\lambda$ is a constant written in terms of $C_{1}$ and $C_{2}$. \medskip

However, for $\beta_{2} = 0$, given the form of Eq.(\ref{Hequation}), one can easily deduce that the resulting solution for the scale factor fails to describe an accelerated expansion ($a(t) \sim t^{\frac{4}{5}}$). We present the equations of reconstruction for the sake of completeness. First, we again consider the form of the model derivatives as
\begin{align}
f_T &= -\frac{1}{12H} \frac{df}{dH}, \\
f_TT &= \frac{1}{144H^2}\frac{d^2 f(H)}{dH^2} - \frac{1}{144H^3}\frac{df(H)}{dH}.
\end{align}

The equation for reconstruction is given by
\begin{align}\label{eqM5b} \nonumber
\frac{d^{2}f(H)}{dH^2} &+ \frac{3(1+\omega)}{\lambda H}\frac{df(H)}{dH} - \frac{3(1+\omega)f(H)}{\lambda H^2}
\nonumber\\
&- \Bigg(\frac{18(1+\omega)}{\lambda} + 12\Bigg) = 0.
\end{align}

A general solution of the Eq.(\ref{eqM5b}) can be written in the form of combinations of power-functions of $H$, i.e., $f(H) \sim \Sigma C_{i} H^{i}$. The coefficients consist of $\omega$, $\lambda$ and integration constants. We here present only a couple of simple examples.   

\begin{enumerate}
\item {For $\omega = 0$,
\begin{equation}\label{slow_var_power}
f(H) = C_{1} H + 6 H^2 + C_{2} H^{-{\frac{3}{\lambda}}}.
\end{equation}
or
\begin{equation}
f(T) = C_{1} \Big(-\frac{T}{6}\Big)^{\frac{1}{2}} - T + C_{2} \Big(-\frac{T}{6}\Big)^{-{\frac{3}{2\lambda}}}.
\end{equation}
}
\item {For $\omega = -1$,
\begin{equation}
f(H) = C_1 + \frac{C_2}{2s}H + 6H^2.
\end{equation}
or
\begin{equation}
f(T) = C_1 + \frac{C_2}{2s}\Big(-\frac{T}{6}\Big)^{\frac{1}{2}} - T.
\end{equation}
}
\end{enumerate}

\begin{figure}
\begin{center}
\includegraphics[width=0.40\textwidth]{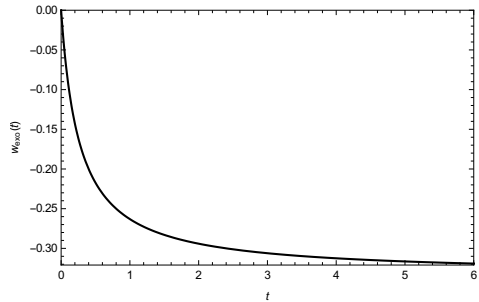}
\caption{Equation of state of the exotic fluid $\omega_{exo}$ as a function of time. $\omega$ for the fluid distribution is chosen to be 0, and plotted for a negative $\lambda$.}
\label{EOS_exotic_variable_jerk_II}
\end{center}
\end{figure}

In Fig.(\ref{EOS_exotic_variable_jerk_II}), the exotic EoS is plotted against time, which tends to a constant negative value far from the necessary $-1$ value to produce a viable model. 

\section{VII. Stability of Reconstructed Models}
In this section, we explore the stability of the models that have been found in this work by taking matter density and Hubble perturbations. In particular, the perturbations are described by (for a similar discussion we refer to the work of Farrugia and Said \cite{said1})
\begin{equation}
H(t) \rightarrow H(t)\left(1+\delta\right), \: \rho(t) \rightarrow \rho(t)\left(1+\delta_{\text{m}}\right),
\end{equation}
where $\delta$ is the deviation of the Hubble parameter $H$ and $\delta_{\text{m}}$ denotes the the deviation of matter over-density. All the deviations are isotropic in nature. These perturbations also infiltrate the arbitrary functional Lagrangian through the expressions
\begin{equation}
\delta f = f_T \delta T, \: \delta f_T = f_{TT} \delta T,
\end{equation}
where $\delta f$ gives the first-order perturbation of $f(T)$ and so on. In this way, it follows that $\delta T = 2T \delta$. Thus, the perturbed forms of the Friedmann equation in Eq.(\ref{eq:00-zero}) and the continuity equation Eq.(\ref{eq:continuity-zero}) become
\begin{align}
-T\left(1+f_T-12H^2f_{TT}\right)\delta &= \kappa^2 \rho \delta_{\text{m}}, \label{eq:00-first} \\
\dot{\delta}_{\text{m}} + 3H(1+w)\delta &= 0, \label{eq:continuity-first}
\end{align}
which govern the evolution of the system. \medskip

$\delta$ and $\delta_{\text{m}}$ are related in terms of $T$, as can be shown from Eq.(\ref{eq:00-zero}) and Eq.(\ref{eq:00-first})
\begin{equation}\label{eq:delta-deltam-rel}
\delta = \dfrac{1}{2T}\dfrac{T+2Tf_T-f}{1+f_T+2Tf_{TT}}\delta_{\text{m}}.
\end{equation}
Therefore from Eq.(\ref{eq:continuity-first}) one can deduce 
\begin{equation}
\dot{\delta}_{\text{m}} + \dfrac{3H}{2T}(1+w)\dfrac{T+2Tf_T-f}{1+f_T+2Tf_{TT}}\delta_{\text{m}}= 0,
\end{equation}
which can be solved to write
\begin{equation}\label{eq:deltam-sol1}
\delta_{\text{m}} = \exp\left[-\dfrac{3}{2}(1+w)\int \dfrac{H}{T}\dfrac{T+2Tf_T-f}{1+f_T+2Tf_{TT}} dt\right],
\end{equation}
which seems to show a dependence on $f(T)$, but this apparent dependence will vanish to reveal the model independent nature of this dependence. \medskip

\begin{figure}
\begin{center}
\includegraphics[width=0.40\textwidth]{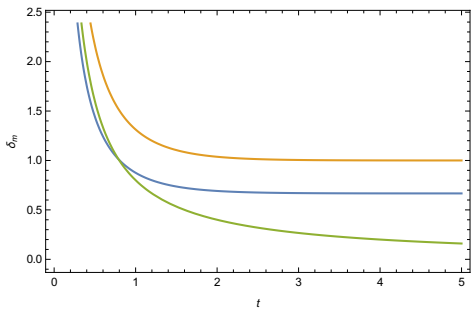}
\caption{Evolution of perturbation of matter overdensity as a function of cosmic time. The blue curve represents the constant jerk parameter model reconstruction. The yellow curve represents the variable jerk parameter model reconstruction where jerk parameter $j \sim \frac{1}{H^2}$. The blue curve shows the evolution for the reconstructed models from a slowly varying jerk.}
\label{delta_m}
\end{center}
\end{figure}

Using the continuity equation Eq.(\ref{eq:continuity-zero}) and the $tt$-component of the field equation Eq.(\ref{eq:00-zero}) with Eq.(\ref{eq:trace-zero}) the integral can be reduced using
\begin{equation}
\dot{H} = \dfrac{1}{4}(1+w) \dfrac{T+2T f_T -f}{1 + f_T + 2T f_{TT}}.
\end{equation}

\begin{figure}
\begin{center}
\includegraphics[width=0.40\textwidth]{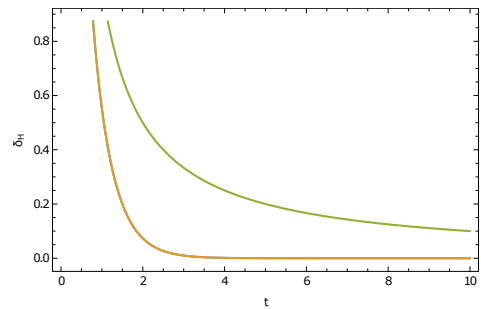}
\caption{Perturbation of the Hubble parameter against cosmic time. The yellow curve represents the evolution for both $(i)$ the constant jerk parameter case and $(ii)$ the jerk proportional to inverse square of hubble parameter case. The green curve represents the evolution for the reconstructed models from a slowly varying jerk.}
\label{delta_H}
\end{center}
\end{figure}

Therefore, from Eq.(\ref{eq:deltam-sol1}) one finds that
\begin{equation}\label{eq:deltam-solution}
\delta_{\text{m}} = \exp\left[\int \dfrac{\dot{H}}{H} dt\right] = \exp\left[\int \dfrac{dH}{H}\right] = kH,
\end{equation}
where $k$ is an integration constant. $k$ can be evaluated from $\delta_{\text{m}}$ evaluated at the present times. Therefore, $k = \delta_{\text{m}}\left(t_0\right)/H_0$. Using Eq.(\ref{eq:delta-deltam-rel}), $\delta$ is determined to be
\begin{equation}
\delta = -\dfrac{\delta_{\text{m}}\left(t_0\right)}{3(1+w)H_0} \dfrac{\dot{H}}{H},
\end{equation}
which is indeed model independent. \medskip

One must note that, the evolution of $\delta$ and $\delta_{\text{m}}$ depend on the current value of $\delta_{\text{m}}$. Moreover, it is easy to see that $w = -1$ gives a singularity for $\delta$. Cosmological stability is present if and only if $\delta_{\text{m}}$ and $\delta$ both are a decreasing function of cosmic time. We investigate different models studied in the present manuscript to comment on the stability. \\

In Fig.(\ref{delta_m}), the evolution of $\delta_m$ is plotted against cosmic time for the various reconstructed models in the preceding sections. While they differ in value, they all tend asymptotically to a constant value which indicates stability. Moreover, in the case of the slowly varying jerk parameter model, the model decays in the same way as the $\Lambda$CDM model. In Fig.(\ref{delta_H}), $\delta$ is shown against cosmic time for all the models under consideration. Again, in this case the models tend to decay with cosmic time.  \\

We also note that the cases involving $\omega = -1$ indicate that there is an absense of matter and the model carries a cosmological constant. This may not be that physical since normal matter is absent and the model contains only gravity. This is well in agreement with the study of cosmological perturbations in $f(T)$ gravity by Chen, Dent, Dutta and Saridakis \cite{chenperturbation}.

\section{VIII. Cosmological Phase Portraits in $f (T)$ cosmology}
We briefly discuss the method of phase portraits in $f(T)$ cosmology before ending the manuscript. The phase space analysis in fact can provide a qualitative idea of the overall cosmological behavior, even if an exact solution can not be found. In the present case, we have found exact solutions in each case depicting a late-time cosmological dynamics. However, we present the phase space analysis anyway, as toy models for some set of arbitrarily chosen parameters, atleast to serve some pedagogical purpose. \\

$f (T)$ gravity allows one to write the torsion scalar $T$ as a function of Hubble $H$. Moreover, from the field equations it is straightforward to see that the physical quantities can also be written in terms of $H$ for example, the torsional energy density, matter energy density. For an equation of state $p = p(\rho)$ assumed at the outset, the pressure is a function of $H$ as well. Thus $\dot{H}$ can be written as
\begin{equation}
\dot{H} = \mathcal{F} (H),
\end{equation}
which is the principle motivation for a phase portrait analysis. This indicates that the cosmological equations can be written as a one-dimensional autonomous system. For details on such a phase space analysis in $f(T)$ theories, we refer to the work of Awad, Hanafy, Nashed and Saridakis \cite{Awad:2017yod}. \\

To work out the application of this dynamical-system analysis for a general $f (T )$ cosmology, one can express the energy density and pressure in term of $H$ from the field equations and write them as

\begin{equation}
\rho = \frac{1}{2\kappa^2} \Big[f(H) - H \frac{df(H)}{dH}\big],
\end{equation}
and
\begin{equation}
p = \frac{1}{6\kappa^2}\dot{H}\frac{d^{2}f(H)}{dH^2} - \rho.
\end{equation}

For a matter distribution with a pre-defined equation of state $p = \omega \rho$, one can then write
\begin{equation}\label{maineq}
\dot{H} = 3 (1 + \omega) \Bigg[ \frac{f(H) - H \frac{df(H)}{dH}}{\frac{d^{2}f(H)}{dH^2}} \Bigg] = \mathcal{F} (H).
\end{equation}

In the present work we do not consider the cases where $\frac{d^{2}f(H)}{dH^2} = 0$. In fact, the $\frac{d^{2}f(H)}{dH^2} = 0$ case corresponds to the trivial $ f(T ) = \alpha \sqrt{T} + \beta$ case, which does not carry any effect of $f(T)$ and the field equations become trivial \cite{caireview}. Eq. (\ref{maineq}) serves as the principle equation of the phase space analysis in $f (T)$ cosmology,pointing out the fixed points or sudden singularities. \\

{\bf Case $1$ : Constant Jerk Parameter} \\

\begin{figure}
\begin{center}
\includegraphics[width=0.40\textwidth]{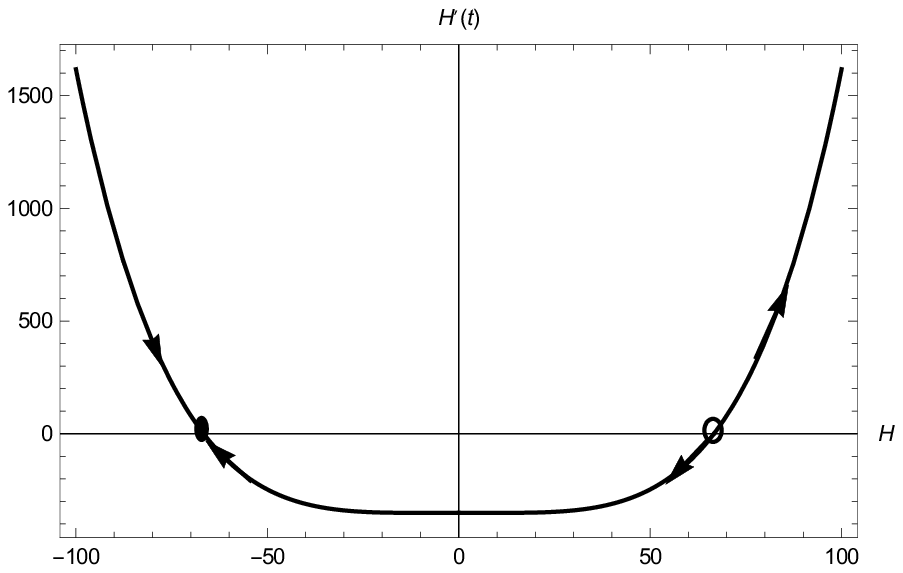}
\includegraphics[width=0.40\textwidth]{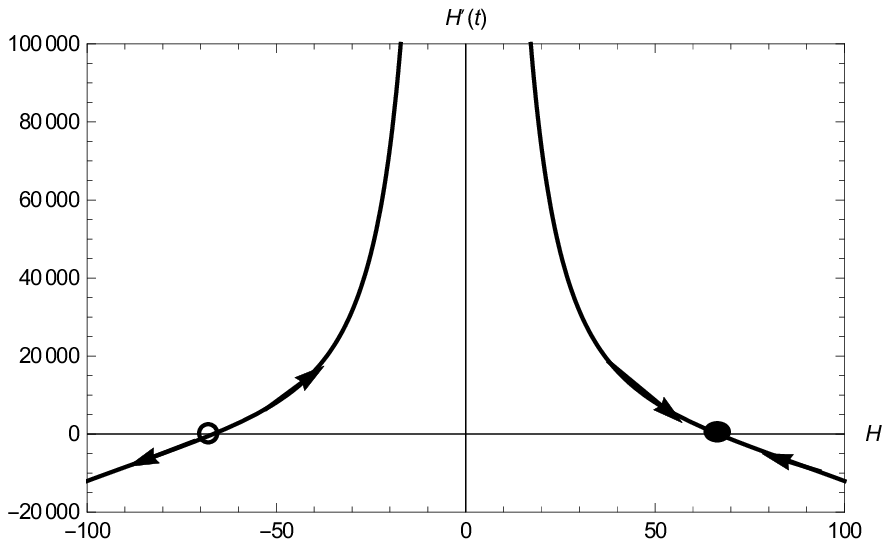}
\includegraphics[width=0.40\textwidth]{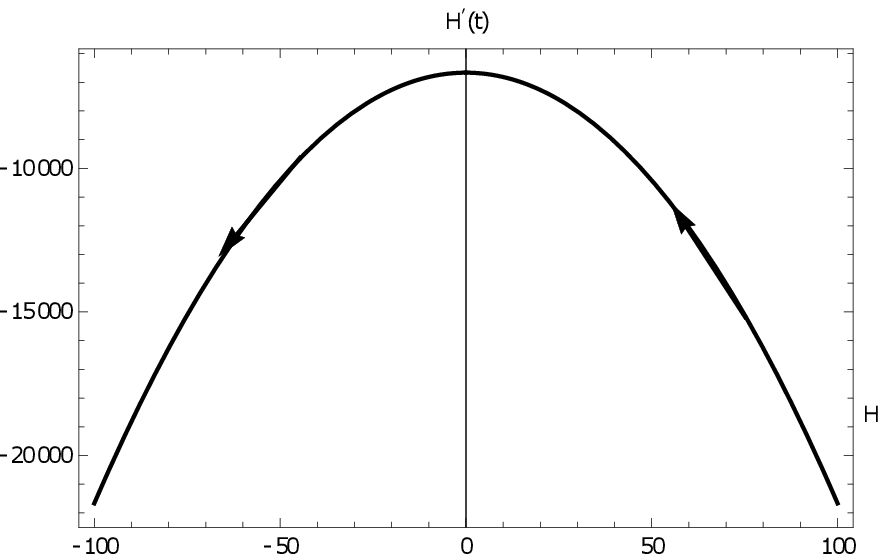}
\includegraphics[width=0.40\textwidth]{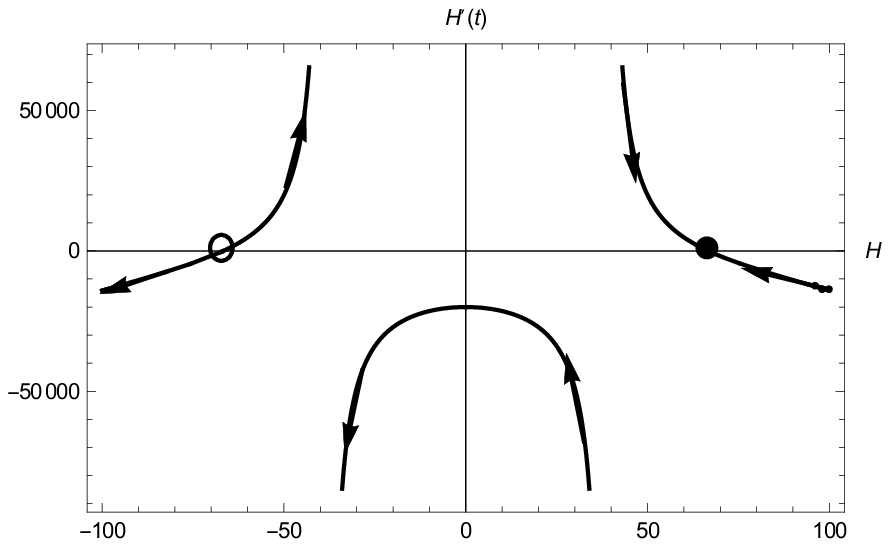}
\caption{Phase plot, or $\dot{H}$ vs $H$ for different values of $\omega$: The figure on the top is for (i) $\omega = -0.9$, $\lambda = 0.01$, $C_{1} = C_{2} = 1$, the second from the top is for (ii) $\omega = 1$, $\lambda = 100$, $C_{1} = C_{2} = 1$, third is for (iii) $\omega = 0$, $\lambda = 100$, $C_{1} = C_{2} = 1$, and bottom plot is for (iv) $\omega = 0.5$, $\lambda = 100$, $C_{1} = C_{2} = 1$.}
\label{Phase_Plot_constant_jerk}
\end{center}
\end{figure}

In Fig. (\ref{Phase_Plot_constant_jerk}), we plot the phase portraits for some choices of the parameters that describe the solution for the Hubble evolution. This allows the different families of solutions to be described in terms of the expansion evolution as trajectories in this phase space. In (i), i.e., for $\omega = -0.9$, a stable de Sitter attractor solution is shown while an eternally accelerating solution is also found. This picture is completely changed for (ii), for $\omega = 1$ where an unstable de Sitter point is present along with a stable $H > 0$ point in this scenario. While (iii)  shows an unrealistic contracting universe for $\omega = 0$, (iv) gives a combination of disjoint evolutions that include both behaviors in (ii) along with an unrealistic contracting universe for $\omega = 0.5$. \\

{\bf Case $2$ : Variable Jerk Parameter $j \sim \frac{1}{H^2}$} \\

\begin{figure}
\begin{center}
\includegraphics[width=0.40\textwidth]{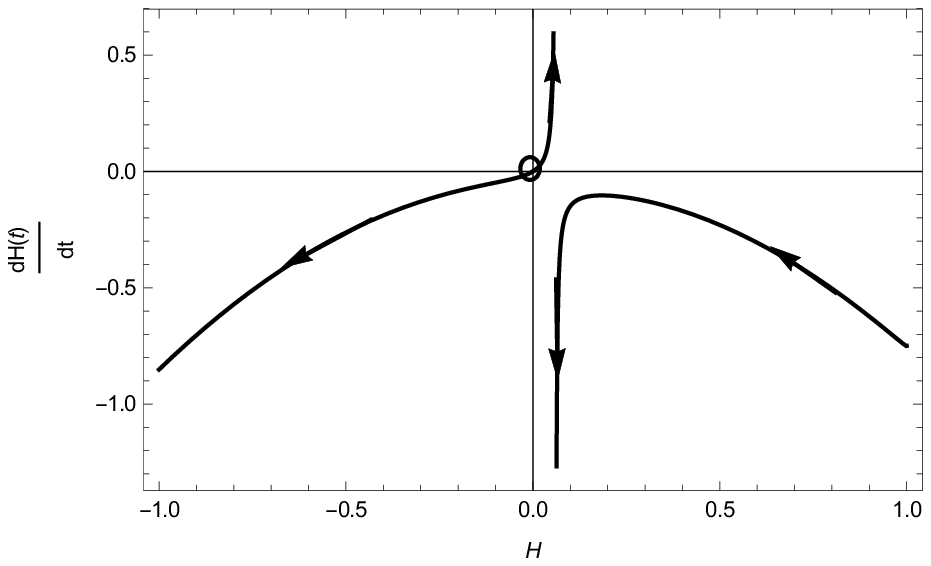}
\includegraphics[width=0.40\textwidth]{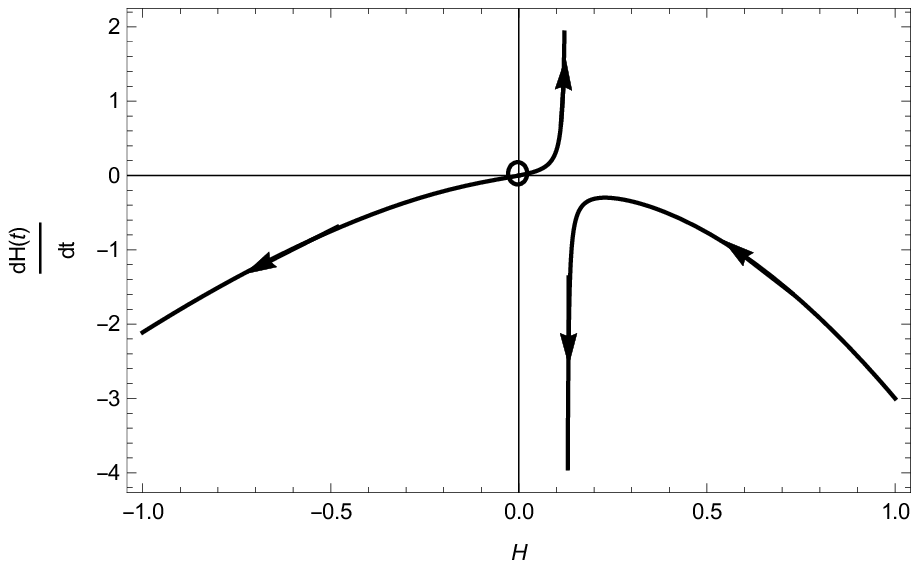}
\includegraphics[width=0.40\textwidth]{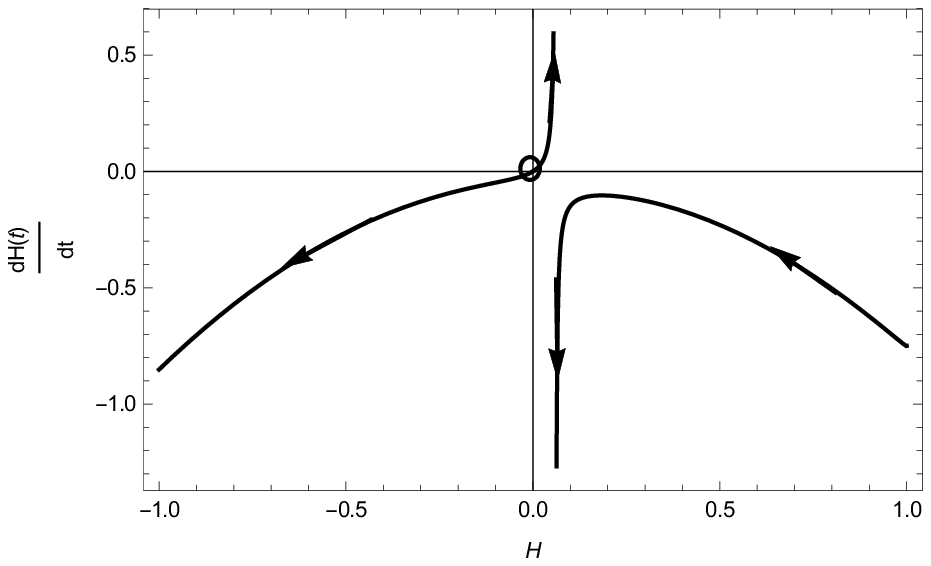}
\includegraphics[width=0.40\textwidth]{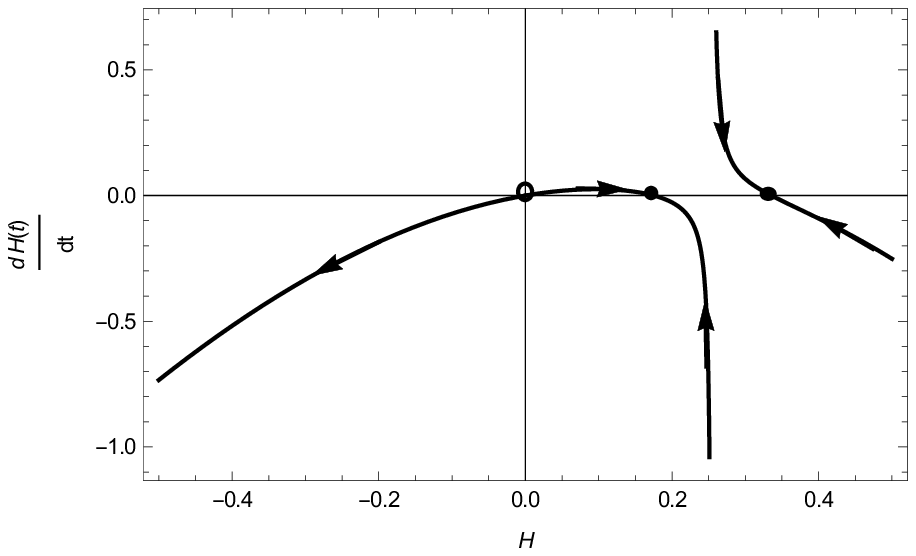}
\caption{Phase plot, or $\dot{H}$ vs $H$ for different values of $\omega$. For all the cases, $C_1 = C_2 = s = 1$ (i). Phase plot for $\omega = 0.$ (ii). Phase plot for $\omega = 1.$ (iii). Phase plot for $\omega = \frac{1}{2}.$ (iv). Phase plot for $\omega = -\frac{1}{2}.$}
\label{phaseplot1}
\end{center}
\end{figure}

\begin{figure}
\begin{center}
\includegraphics[width=0.45\textwidth]{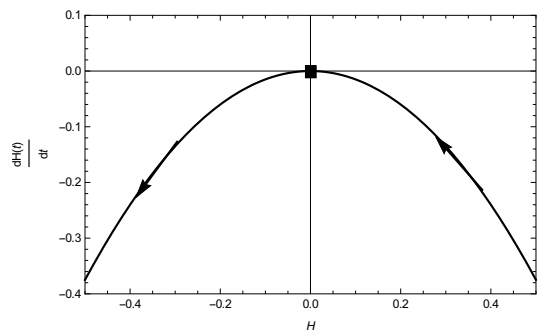}
\caption{Phase plot for $C_1 = C_2 = 0$. $s = 1$.}
\label{phaseplot4}
\end{center}
\end{figure}

In Fig.(\ref{phaseplot1}), plots (i,ii,iii) show a behavior where an unstable Minkowski fixed point is accompanied with an eternally contracting Universe. In (iv), the flow of the unstable Minkowski point has a second fixed point which is an attractor de Sitter point, while the other evolution turns out to have a fixed point that is stable. In Fig.(\ref{phaseplot4}), the special case of $C_1=0=C_2$ is explored where a semi-stable Minkowski point is found.  \\

{\bf Case $3$ : Slowly Varying Jerk Parameter} \\

\begin{figure}
\begin{center}
\includegraphics[width=0.40\textwidth]{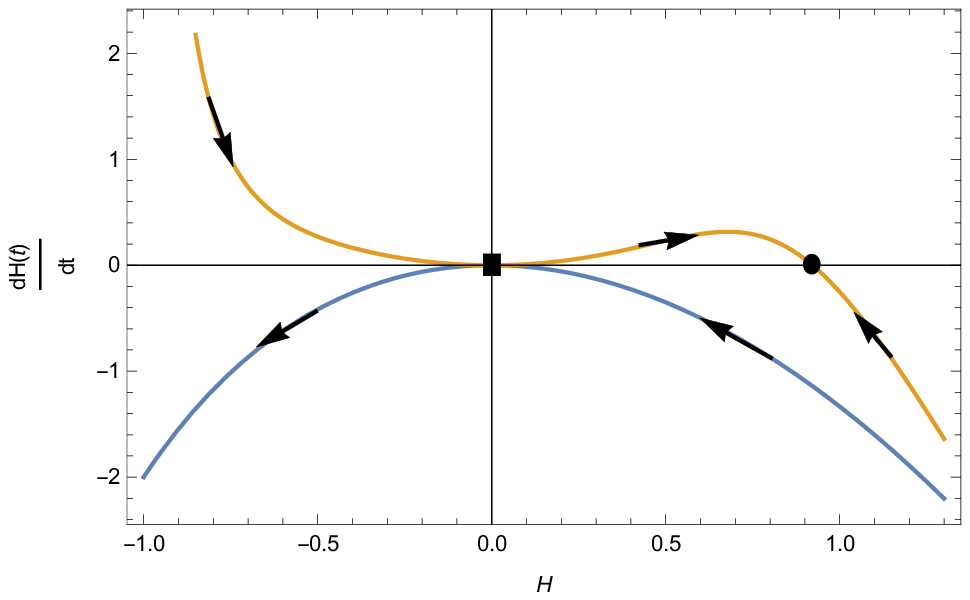}
\caption{Phase plot for $\omega = 0$. The yellow curve is for $\lambda = 1$ and the blue curve is for $\lambda = -1$. $C_{1}$ and $C_{2}$ are chosen to be unity.}
\label{phaseplot5}
\end{center}
\end{figure}

\begin{figure}
\begin{center}
\includegraphics[width=0.40\textwidth]{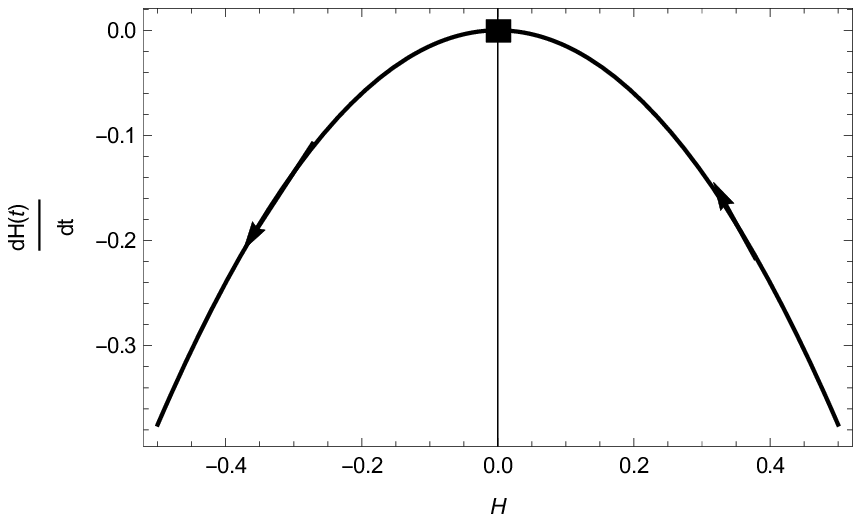}
\caption{Phase plot for $\omega = -1$. $C_{1}$ and $C_{2}$ are chosen to be unity.}
\label{phaseplot6}
\end{center}
\end{figure}

 The phase plots are explored for two separate EoS scenarios, where in the first instance (Fig. \ref{phaseplot5}) $\omega = 0$ and two behaviors follow one of which has a stable de Sitter point; in the second scenario (Fig. \ref{phaseplot6}) $\omega = -1$ and a semi-stable Minkowski fixed point follows.

\section{IX Conclusion}
In the present work, the possibility of using the cosmological jerk parameter to reconstruct the modified Teleparrallel lagrangian has been explored analytically. This parrallel avenue of torsion based modification of gravity is experiencing a fast growing interest amongst the community with a particular focus on cosmological models, given the recent results by the Planck Collaboration \cite{Aghanim:2018eyx}. While teleparallel gravity reproduces GR in some limit, it is not equivalent in terms of other behavior and may solve some of the outstanding theoretical issues of GR. \medskip

Three different analytical models have been considered in this work which present a scheme of reverse engineering the cosmological parameters from a choice of the jerk parameter. The constant jerk parameter case is inspired by the $\Lambda$CDM model and produces a more general setup when compared with the TEGR Lagrangian. This limits to TEGR for a particular choice of parameters but can produce a much more general parameter space for comparison against data in further work. The second model presupposes a particular dependence on the Hubble parameter such that there is an inverse square relation dependence. While the system is intractable in general, it can be solved for particular choices of the matter EoS parameter with an interesting exotic EoS parameter profile which can produces models not dissimilar to current observations. The Hubble free luminosity distance data is fitted with the corresponding observational data to find a good fit. Finally, in the third model, we consider a slowly varying jerk parameter, which are certainly applicable in late-time cosmology. The system of equation in this case turns out to be difficult to solve in general and we solve for a very particular set of parameters ($\beta_{2} = 0$). Although this instance does not actually inspire a late time accelerating solution, a general analysis ($\beta_{2} \neq 0$) of the slowly varying jerk parameter case may produce more significant solutions and will be a part of future work. \medskip

The present work is basically a reconstruction of theoretical cosmology from the kinematical variables only, without knowing any apriori expansion history. In the present case the kinematic variable being the jerk parameter, higher order parameters (such as snap parameter, involving $a''''(t)$) can be used for a more general reconstruction scheme as well, provided the mathematics can be addressed. A reconstruction from kinematic quantity alone is exciting in the sense that it can in principle also be applied in order to reconstruct $f(T)$ for a jerk parameter that corresponds to a bounce behavior \cite{bounce}. Cosmological reconstruction offers an interesting store of possibilities, providing new models of gravity. The models presented in the manuscript are simple, realistic and easy to work with for further investigations. However, we note here that the models only present some special cases while the general picture may consist of far more plausible set of solutions for a completel general jerk parameter.

\section{Acknowledgements}
SC was supported by the National Post-Doctoral Fellowship (file number: PDF/2017/000750) from the Science and Engineering Research Board (SERB), Government of India. The authors would like to acknowledge networking support by the COST Action GWverse CA16104. This article is based upon work from CANTATA COST (European Cooperation in Science and Technology) action CA15117, EU Framework Programme Horizon 2020. The work of KB was supported in part by the JSPS KAKENHI Grant Number JP25800136 and Competitive Research Funds for Fukushima University Faculty (18RI009).


\begin{thebibliography}{99}
\bibitem{obs} S. Perlmutter et al. [SNCP Collaboration], Astrophys. J. {\bf 517}, 565 (1999); A. G. Riess et al. [SNST Collaboration], Astron. J. {\bf 116}, 1009 (1998); D. N. Spergel et al. [WMAP Collaboration], Astrophys. J. Suppl. {\bf 148}, 175 (2003); ibid. {\bf 170}, 377 (2007); E. Komatsu et al. [WMAP Collaboration], Astrophys. J. Suppl. {\bf 180}, 330 (2009); E. Komatsu et al., arXiv:1001.4538 [astro-ph.CO]; M. Tegmark et al., Phys. Rev. D. {\bf 69}, 103501 (2004); U. Seljak et al. [SDSS Collaboration],
Phys. Rev. D. {\bf 71}, 103515 (2005); D. J. Eisenstein et al., Astrophys. J. {\bf 633}, 560 (2005); B. Jain and A. Taylor, Phys. Rev. Lett. {\bf 91}, 141302 (2003).

\bibitem{Akrami:2018odb} Y.~Akrami {\it et al.} [Planck Collaboration], arXiv:1807.06211 [astro-ph.CO].

\bibitem{Aghanim:2018eyx} N. Aghanim {\it et al.} [Planck Collaboration], arXiv:1807.06209 [astro-ph.CO].
  
\bibitem{Peter:2012rz} A. H. G. Peter, arXiv:1201.3942 [astro-ph.CO].

\bibitem{CST} E. J. Copeland, M. Sami and S. Tsujikawa, Int. J. Mod. Phys. D. {\bf 15}, 1753 (2006).

\bibitem{weinberg} S. Weinberg, Rev. Mod. Phys. {\bf 61} 1 (1989).

\bibitem{scalarfield1} B. Ratra and P. J. E. Peebles, Phys. Rev. D. {\bf 37}, 3406 (1988).

\bibitem{scalarfield2} J. M. Overduin and F. I. Cooperstock, Phys. Rev. D. {\bf 58}, 043506 (1998).

\bibitem{scalarfield3} A. W. Brookfield, C. van de Bruck, D. F. Mota and D. Tocchini-Valentini, Phys. Rev. Lett. {\bf 96}, 061301 (2006).

\bibitem{scalarfield4} M. C. Bento, O. Bertolami and A. A. Sen, Phys. Rev. D. {\bf 66}, 043507 (2002).

\bibitem{Aldrovandi:2013wha} R. Aldrovandi and J. G. Pereira, Fundam.\ Theor.\ Phys.\  {\bf 173} (2013).

\bibitem{caireview} Y-F Cai, S. Capozziello, M. DeLaurentis and E. N. Saridakis, Rept. Prog. Phys. {\bf 79}, 106901 (2016).

\bibitem{telep1} F. W. Hehl, P. Von Der Heyde, G. D. Kerlick and J. M. Nester, Rev. Mod. Phys. {\bf 48}, 393 (1976).

\bibitem{telep2} K. Hayashi and T. Shirafuji, Phys. Rev. D. {\bf 19}, 3524 (1979).

\bibitem{telep3} E. E. Flanagan and E. Rosenthal, Phys. Rev. D. {\bf 75}, 124016 (2007).

\bibitem{telep4} A. Einstein, Sitzungsber. Preuss. Akad. Wiss. Phys. Math. Kl., {\bf 217} (1928); {\bf 401} (1930); A. Einstein, Math. Ann. {\bf 102}, 685 (1930).

\bibitem{krssak} M. Krssak and E. N. Saridakis, Class. Quant. Grav. {\bf 33} 115009 (2016).

\bibitem{Nojiri:2010wj} S. Nojiri and S. D. Odintsov, Phys. Rept. {\bf 505}, 59 (2011).

\bibitem{Capozziello:2011et} S. Capozziello and M. De Laurentis, Phys. Rept. {\bf 509}, 167 (2011).

\bibitem{Capozziello:2010zz} V. Faraoni and S. Capozziello, Fundam. Theor. Phys. {\bf 170}, (2010).

\bibitem{Bamba:2015uma} K. Bamba and S. D. Odintsov, Symmetry {\bf 7}, 220 (2015).
  
\bibitem{Nojiri:2017ncd} S. Nojiri, S. D. Odintsov and V. K. Oikonomou, Phys. Rept. {\bf 692}, 1 (2017).
  
\bibitem{Bamba:2012cp} K. Bamba, S. Capozziello, S. Nojiri and S. D. Odintsov, Astrophys. Space Sci. {\bf 342}, 155 (2012).

\bibitem{Farrugia:2018gyz} G. Farrugia, J. L. Said, V. Gakis and E. N. Saridakis, Phys.\ Rev.\ D {\bf 97} no.12, 124064 (2018).

\bibitem{ben} G. Bengochea and R. Ferraro, Phys. Rev. D. {\bf 79} : 124019, (2009).

\bibitem{Linder:2010py} E. V. Linder, Phys.\ Rev.\ D. {\bf 81} : 127301, (2010); Erratum: [Phys.\ Rev.\ D {\bf 82} (2010) 109902].

\bibitem{wuyu1} P. Wu and H. Yu, Phys. Lett. B. {\bf 692}, 176 (2010).

\bibitem{wuyu2} P. Wu and H. Yu, Phys. Lett. B. {\bf 693}, 415 (2010).

\bibitem{capo1} S. Capozziello, V. F. Cardone, H. Farajollahi and A. Ravanpak, Phys. Rev. D. {\bf 84} :043527, (2011).

\bibitem{Nunes:2018evm} R. C. Nunes, S. Pan and E. N. Saridakis, Phys.\ Rev.\ D. {\bf 98}, no.10, 104055 (2018).

\bibitem{bamba1} K. Bamba, C-Q. Geng, C-C. Lee and L-W. Luo, JCAP {\bf 1101} : 021, (2011).

\bibitem{behboodi} A. Behboodi, S. Akhshabi and K. Nozari, Phys. Lett. B. {\bf 718}, 30 (2012).

\bibitem{palia} A. Paliathanasis, J. D. Barrow and P.G.L. Leach, Phys. Rev. D. {\bf 94}, 023525 (2016).

\bibitem{Paliathanasis:2017htk} A. Paliathanasis, J. L. Said and J. D. Barrow, Phys.\ Rev.\ D. {\bf 97}, no.4, 044008 (2018).

\bibitem{Farrugia:2016qqe} G. Farrugia and J. L. Said, Phys.\ Rev.\ D. {\bf 94}, no.12, 124054 (2016). 



\bibitem{ruggiero} M. L. Ruggiero and N. Radicella, Phys. Rev. D. {\bf 91}, 104014 (2015).

\bibitem{Finch:2018gkh} A. Finch and J. L. Said, Eur.\ Phys.\ J.\ C. {\bf 78}, no.7, 560 (2018). 

\bibitem{said} G. Farrugia, J. L. Said and M. L. Ruggiero, Phys. Rev. D. {\bf 93}, 104034 (2016).

\bibitem{said1} G. Farrugia and J. L. Said, Phys. Rev. D. {\bf 94}, 124054 (2016).

\bibitem{nojireco1} S. Nojiri and S. D. Odintsov, Phys. Rev. D. {\bf 74}, 086005 (2006).

\bibitem{nojireco2} S. Nojiri and S. D. Odintsov, J. Phys. Conf. Ser. {\bf 66}, 012005 (2007).

\bibitem{nojireco3} S. Nojiri and S. D. Odintsov, Phys. Lett. B. {\bf 576}, 5 (2003).

\bibitem{myrza1} R. Myrzakulov, Eur. Phys. J. C. {\bf 71} : 1752, (2011).

\bibitem{dent} J. B. Dent, S. Dutta and E. N. Saridakis, JCAP {\bf 1101} : 009, (2011).

\bibitem{myrza2} R. Myrzakulov,	Gen. Relativ. Gravit. {\bf 44}, 3059 (2012).

\bibitem{bamba2} K. Bamba, R. Myrzakulov, S. Nojiri and S. D. Odintsov, Phys. Rev. D. {\bf 85}, 104036 (2012).

\bibitem{scjs} S. Chakrabarti, J. L. Said and G. Farrugia, Eur. Phys. J. C. {\bf 77} : 815 (2017).

\bibitem{scjskbfR} S. Chakrabarti, J. L. Said and K. Bamba, arXiv:1809.04955v2 [gr-qc].  

\bibitem{ankan} A. Mukherjee and N. Banerjee, Phys. Rev. D. {\bf 93}, 043002 (2016).

\bibitem{nojiodisaez} S. Nojiri, S. D. Odintsov and D. Saez-Gomez, Phys. Lett. B. {\bf 681} : 74 (2009).

\bibitem{chenperturbation} S. Chen, J. B. Dent, S. Dutta and E. N. Saridakis, Phys. Rev. D. {\bf 83} :023508 (2011).

\bibitem{Awad:2017yod} A.~Awad, W.~El Hanafy, G.~G.~L.~Nashed and E.~N.~Saridakis, JCAP, {\bf 1802} no.02,  052 (2018).  
    
\bibitem{Carroll:2003wy} S.~M.~Carroll, V.~Duvvuri, M.~Trodden and M.~S.~Turner, Phys. Rev. D. {\bf 70} : 043528 (2004).

\bibitem{bounce} Y. F. Cai, S. H. Chen, J. B. Dent, S. Dutta and E. N. Saridakis, Class. Quant. Grav. {\bf 28} 215011 (2011).

\end{thebibliography}
\end{document}